\documentclass{article}

\pdfoutput=1

\usepackage{graphicx}
\usepackage{amsmath,amssymb,bm}
\usepackage[noblocks]{authblk}
\catcode`\_=\active
\newcommand_[1]{\ensuremath{\sb{\mathrm{#1}}}}
\author[a,b]{Alberto Giacomello\thanks{giacomello@is.mpg.de}}
\author[a]{Lothar Schimmele}
\author[a,c]{Siegfried Dietrich}

\affil[a]{Max-Planck-Institut f\"ur Intelligente Systeme, Stuttgart, Germany}
\affil[b]{Dipartimento di Ingegneria Meccanica e Aerospaziale, Sapienza Universit\`a di Roma, Rome, Italy}
\affil[c]{IV. Institut f\"ur Theoretische Physik, Universit\"at Stuttgart, Stuttgart, Germany}

\date{}

\title{Wetting hysteresis induced by nanodefects}

\begin{document}
	%NB The command underscore "_" has been redefined to ensure that all
	%subscripts in math mode are in roman fonts. Avoid underscores in
	%figure names.

%\doublespacing
\maketitle 

\begin{abstract}
Wetting of actual surfaces involves diverse hysteretic phenomena
stemming from ever-present imperfections.
Here we clarify the origin of wetting hysteresis for a liquid front
advancing or receding across an isolated defect of nanometric size.
Various kinds of chemical and topographical nanodefects are investigated
which represent salient features of actual heterogeneous surfaces.  
The most probable wetting path across surface heterogeneities is identified
by combining, within an innovative approach, microscopic classical density functional
theory and the string method devised for the study of rare events. 
The computed rugged free energy landscape demonstrates that hysteresis 
emerges as a consequence of metastable pinning of the liquid front at
the defects; the barriers for thermally activated defect crossing, the pinning force, 
and hysteresis are quantified and related to the geometry and chemistry of
the defects allowing for the occurrence of nanoscopic effects. 
The main result of our calculations is that even weak nanoscale
defects, which are difficult to characterize in generic microfluidic
experiments, can be the source of a plethora of hysteretical phenomena,
including the pinning of nanobubbles.

\vspace{0.5cm}
\paragraph{Significance Statement}
Drops may fail to slide even on extremely smooth surfaces.  This
fact is due to the ubiquitous imperfections of surfaces:  defects of
molecular size --  here too small to be noticed and experimentally characterizable --  can
hinder a macroscopic drop.
In order to understand these far-reaching
nanoscale phenomena we combine a molecular description of the liquid
with advanced techniques for the study of rare events.  This approach
allows us to bridge the diverse scales involved, demonstrating that
even weak nanometer-sized surface defects can give rise to measurable differences
between advancing and receding liquid fronts, i.e., contact angle
hysteresis. The present results also shed new light on the unexpectedly
long life of surface nanobubbles.

\end{abstract}

Within the burgeoning field of wetting \cite{bonn2009,dietrich1988}, the behavior of droplets on
heterogeneous surfaces is one of the most active subjects of research
\cite{kusumaatmaja2007,savva2013,musterd2014,dubov2014,berim2015,wang2015}. 
Indeed, the interaction of a liquid front with surface heterogeneities
poses challenges concerning even the static behavior
which are still open nowadays \cite{bonn2009}. Here we
take a fresh look at this
subject using microscopic density functional theory in
conjunction with
the string method for the study of rare events.

More specifically,
the present study deals with a liquid front which advances or recedes
over a surface with isolated nanometer-sized heterogeneities.  All
surfaces of practical use do feature heterogeneities at this scale,
forming either chemical blemishes or topographical defects such as
bumps or cavities; even advanced fabrication techniques cannot
completely prevent them to occur.  Although tiny and thus difficult to
control and to characterize, these nanoscale defects have remarkable
\emph{macroscopic} consequences, such as generating measurable differences
between advancing and receding liquid fronts -- i.e., the
\emph{contact angle hysteresis}
investigated here.

The phenomena which originate from the interaction between surface
heterogeneities and liquids are diverse and various
names are used to underscore different aspects. For instance,
``pinning'' refers to hindrance of the motion of the contact line,
``wetting hysteresis'' is here defined as the qualitative
difference between advancing and receding processes,
while ``contact angle hysteresis'' indicates the difference observed between the
apparent contact angle of advancing and receding drops. 
Understanding and relating all these different aspects of
the problem requires
a thorough theoretical investigation about the origin of hysteresis in wetting:
many years after the first systematic studies 
\cite{johnson1964,dettre1964}, contact angle hysteresis continues to be
a very active research topic in physics, physical chemistry, and
materials science \cite{bonn2009}.

The issue of surface nanobubbles \cite{tyrrell2001,lohse2015a} is a contemporary example of the debate
on wetting hysteresis. The pinning of their contact line seems to be the
crucial element for explaining the unexpectedly long lifetime of
nanobubbles. Recent findings have shown that nanobubbles can
survive for hours due to a combination of contact line pinning and
retarded one-dimensional diffusion \cite{weijs2013,liu2014,lohse2015}. Zhang \emph{et al.}
experimentally verified that nanobubbles are indeed pinned at the
surface and that their apparent shape is that of a spherical cap 
\cite{zhang2013}. The size of surface nanobubbles, with a typical base
radius $< 1\; \mu m$, and the regularity of their shape suggest that the
pinning originates from much smaller surface defects, \emph{i.e.}, in the
nanometer range. However, it is still an open question whether defects
below $100$~nm can pin a liquid-gas interface at
all \cite{johnson1964,huh1977} and thus whether they can be the cause of
the stability of nanobubbles.

Recent experiments, using atomic force microscopy, reported the pinning
force exerted by nanoscale defects \cite{ondarcuhu2005,delmas2011}:
Ondar{\c{c}}uhu \emph{et al.} were able to identify single pinning or
depinning events.  However, to the best of our knowledge there are no
wetting experiments, involving a single well-characterized nanoscale
defect, which are capable to relate the characteristics 
of the defect with the pinning
force. Therefore, it is crucial to resolve how the chemical and
topographical nano-features of surfaces determine
contact angle hysteresis.
Another intriguing and still open question is \emph{how} a liquid
front advances or recedes across such features.

Pinning of the contact line is important not only for static contact
angle hysteresis but also for contact line motion (see, e.g.,
Ref.~\cite{chibbaro2009}).  In the case of
nanoscale defects, experiments suggest that the mechanism of contact
line motion is governed by thermally activated pinning and depinning
events \cite{rolley2007,ramiasa2013}. Indeed,  the ``pinning energy'' (\emph{i.e.}, the free energy barrier)
related to nanodefects may be of the order of $k_B T$ \cite{delmas2011};
thus, depending on the physical and chemical properties of the
nanodefects, the migration across a defect may be a deterministic
process driven by an external force or a stochastic one induced by
thermal fluctuation. 
Recent experiments on colloids at the air-water interface suggest that
thermal activation of local deformations of the triple line, possibly induced by
pinning at nanodefects, can explain their anomalous diffusion
coefficient \cite{boniello2015}.
These experimental evidences also call for a better
characterization of the advance and the retreat of liquids over nanodefects
in terms of free energy barriers, revealing the type of mechanism at work and 
the associated rate of the process. 

Contact angle hysteresis was studied theoretically
in the seminal paper by Joanny and de Gennes alluding to strong,
isolated defects with characteristic size larger than ca. $30$~nm \cite{joanny1984}. 
Therein the macroscopic deformation of
the liquid-vapor interface was worked out by assigning a given defect
``force'' both to smooth and to mesa defects having sharp discontinuities.
Robbins and Joanny \cite{robbins1987} focused on weak, macroscopic heterogeneities
which individually are incapable to pin the interface but 
which induce contact angle hysteresis \emph{via} collective pinning by many
defects. Yeomans \emph{et al.} performed mesoscale ($>50$~nm) lattice
Boltzmann simulations for drops and liquid slabs in channels in order 
to calculate contact angle hysteresis on
superhydrophobic surfaces and dense chemical patterns 
\cite{kusumaatmaja2007,mognetti2010}. This way, a quasi-static numerical experiment
was performed, which determines the (meta)stable configurations at given drop
radii \cite{kusumaatmaja2007} or at prescribed apparent contact angles 
\cite{mognetti2010}.  A similar approach was followed by Semprebon
\emph{et al.} \cite{semprebon2012} who used Surface Evolver \cite{brakke1992}  in order 
to compute the modes of advance of the liquid on macroscopically
structured surfaces. 
The string method was used in connection with Ginzburg-Landau models in order to
explore the Cassie-Wenzel transition  for droplets on
superhydrophobic surfaces \cite{zhang2014,pashos2015}.
To date only few studies investigated how the ``pinning force'' or the
corresponding free
energy barriers are related to the geometrical and chemical characteristics of
defects \cite{kusumaatmaja2007,liu2014}. Furthermore, at the nanoscale,
at which the typical dimensions of the roughness are comparable to the size of the
liquid particles and where van der Waals forces come into play,
wetting hysteresis as well as advancing and receding mechanisms 
remain still unexplored.

The present study aims at a microscopic understanding of the mechanism
of hysteresis induced by chemical or topographical defects at the
nanoscale. Contact line motion at that scale might also be a thermally
assisted process \cite{rolley2007,ramiasa2013}. We therefore discuss,
within the same framework, both macroscopic deterministic pinning or
depinning, as has been done in many previous analyses, as well as thermally
activated depinning. Predicting thermally activated pinning or depinning rates
requires \emph{inter alia} the knowledge of the corresponding activation barriers.  
Thus, in order to understand hysteresis one has to go beyond the equilibrium
picture and investigate how the free energy landscape,
including features such as metastable minima and activation barriers,
is influenced by deterministic forces driving the liquid.  
Here we have combined a microscopic description of the
fluid in terms of the classical density functional theory (DFT) due to
Rosenfeld, which is reliable on molecular scales
\cite{rosenfeld1989,rosenfeld1997,roth2010}, with
the string method \cite{e2002}, a state-of-the-art tool for the study of
rare events. This facilitates to accurately
characterize the mechanism and the free energy barriers associated with a
liquid wedge advancing or receding over a single chemical or
topographical nanodefect.  
The string method allowed us to overcome the
limitations of brute force calculations in the presence of
metastabilities, to identify the \emph{transition path} 
\footnote{Here we use the notion ``most probable path'' or,
more concisely, ``transition path'' in order to denote the most probable succession of
configurations which the liquid assumes while it advances or recedes
past a heterogeneity.} (shown in the videos in the \emph{Supporting
Information}), and to provide accurate estimates for free energy barriers
for wetting or drying across a single heterogeneity.  Since we are
including microscopic details of the fluid, our approach can capture the
diverse effects related to nanodefects. As demonstrated by our
results, for such defects predictions based on macroscopic concepts like surface energies and
contact angle become unreliable and incomplete. 
Given the experimental difficulty to access simultaneously the
geometry and the chemistry of a nanodefect and the force exerted by it on
advancing or receding liquid fronts, the present calculations are the
first attempt to fully characterize nanoscale wetting hysteresis. 

The paper is organized such that the first section is devoted to the
description of the computational methods, with particular emphasis on 
novel aspects. The second section is concerned with the results on
wetting hysteresis on single defects and the discussion of them. The third section 
investigates the role of dilute distributions of nanodefects on the macroscopic 
advancing and receding contact angles (i.e., contact angle hysteresis).

%%%%%%%%%%%%%%%%%%%%%%%%%%%%%%%%%%%%%%%%%%%%%%%%%%%%%%%%%%%%%%%%%%%%%

\section{Combined DFT and string calculations}

%%%%%%%%%%%%%%%%%%%%%%%%%%%%%%%%%%%%%%%%%%%%%%%%%%%%%%%%%%%%%%%%%%%%%

\subsection{Fundamental measure theory}

Classical density functional theory (DFT)
allows one to determine the equilibrium number density of a fluid by
minimizing the grand potential functional $\Omega$ of the system under
investigation:
\begin{equation}
	\Omega[\rho] = F[\rho] + \int \;\mathrm{d}^3 r \; \rho(\bm r) [V(\bm r)
	-	\mu] \text{ ,}
	\label{eq:omega}
\end{equation}
where $F[\rho]$ is the intrinsic Helmholtz free energy functional
encompassing the fluid-fluid interactions, $\rho(\bm r)$
is the fluid number density at position $\bm r$, 
$V(\bm r)$ is the substrate potential, and $\mu$
is the chemical potential.

For $F[\rho]$ we have chosen the fundamental measure theory of Rosenfeld
\cite{rosenfeld1989,rosenfeld1997,roth2010}, which accurately accounts
for the repulsive part of the interaction between the fluid particles.
For the attractive part of the fluid-fluid and fluid-solid interaction
we have adopted forces of the van der Waals type.
Their length and energy
scales are set by the particle diameter $\sigma$ and by
the interaction strength $\varepsilon$.
The attractive part of the wall-fluid interaction is expressed in terms of 
the wall energy $u_{w}$,
which determines the attractive potential $V_{\mathrm{att}}(d) =
-u_{w}/d^3$ acting on a fluid particle at a distance $d$ from the planar
surface of a half-space filled continuously and homogeneously
by the same particles as the wall. 
In order to mimic  chemical or topographical surface heterogeneities 
the wall energy $u_{w}(\bm r)$ can vary spatially.  Concerning the
detailed forms of
$F[\rho]$ and $V(\bm r)$ as well as the calculation details see
Refs.~\cite{roth2010,singh2015} and the \emph{Supporting Information}.

\subsection{Model system}

\begin{figure}[t!]
	\centering
	\includegraphics[width=0.8\textwidth]{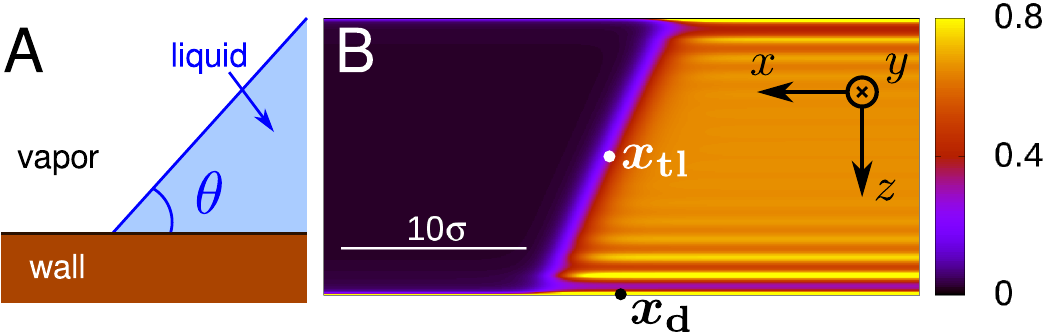}
	\caption{\emph{A}) Macroscopic model (solid line) of a liquid wedge
		meeting the solid wall with an angle $\theta=\theta_Y\equiv
		\theta_Y^{\mathrm{low}}$.  \emph{B}) Actual liquid ``wedge'' used in
		DFT calculations.  The color code indicates $\rho(\bm r)\sigma^3$.
		The position $x_{tl}$ of the triple line in $x$ direction is marked
		by the white dot, typically close to the middle of the liquid-vapor
		interface.  The defects introduced in the next section (not present
		here) will be placed at the center of the lower wall at position
		$x_{d}$ along the x axis (black dot) in order to ensure that the
		boundary conditions on the left and right do not interfere with the
		migration process of the liquid wedge across the
	defects.\label{fig:system}}
\end{figure}

We consider a liquid wedge defined as a semi-infinite liquid domain bounded by a
planar liquid-vapor interface which forms an angle $\theta$ with the
substrate (see Fig.~\ref{fig:system}A).  In a macroscopic description
of the liquid wedge, this angle coincides with the equilibrium Young 
contact angle $\cos\theta_Y
\equiv(\gamma_{sv}-\gamma_{sl})/\gamma_{\,lv}$, where $\gamma_{sv}$, 
$\gamma_{sl}$, and $\gamma_{\,lv}$ denote the
surface tensions of the solid-vapor, solid-liquid, and liquid-vapor
interfaces, respectively. 

%Fig1

The system used in our DFT calculations for producing the liquid wedge
in a finite computational box consists of two planar, parallel walls with the wall
energies $u_w^{\mathrm{up}}$ and $u_w^{\mathrm{low}}$ tuned such 
that the liquid-vapor interface meets the two walls with
complementary macroscopic angles $\theta_Y^{\mathrm{low}}$ and 
$\theta_Y^{\mathrm{up}}=180^\circ -\theta_Y^{\mathrm{low}}$
(see Fig.~\ref{fig:system}B).  Specifically, the lower wall is slightly
lyophilic with $u_w^{\mathrm{low}}=3\varepsilon$ corresponding to
$\theta_Y^{\mathrm{low}}=80^\circ$ while the upper wall exhibits
$u_w^{\mathrm{up}}=2.6\varepsilon$ corresponding to
$\theta_Y^{\mathrm{up}}=100^\circ$.  The angles correspond to our choice $k_BT =
\varepsilon$ for the temperature $T$; the chemical potential is that for
liquid-vapor coexistence in the bulk, $\mu=\mu_0(T)$, so that 
$\rho_l \sigma^3=0.658$ and $\rho_v \sigma^3=0.0199$, respectively.
The computational box size is $L_x \times L_y \times L_z = 
32.5\sigma \times 17.5\sigma \times 16\sigma$

In Fig.~\ref{fig:system}B we show the equilibrium profile for the fluid
number density in the presence of defect-free, planar walls. It reproduces
the salient features of a liquid wedge including, near the wall the
layering on the liquid side and the wetting film formation on the vapor
side (see also Ref.~\cite{nold2014}).  This defect-free system is also
used as a reference system in order to subtract small background variations of the
free energy due to translations of the liquid wedge which result from
the shift of the capillary liquid-vapor coexistence \cite{evans1986}
relative to bulk coexistence due to various finite size effects (see
Fig.~S5).

The system used to study wetting hysteresis in the next section differs from
the system in Fig.~\ref{fig:system} in that the lower wall features a
single defect.  
We consider the following defects
(for details, see the \emph{Supporting Information}):
\begin{itemize}
	\item chemical defects: lyophobic ($u_w=2\varepsilon$),
		partially wet ($u_w=3.5\varepsilon$), and
		completely wet ($u_w=4\varepsilon$) $\Delta x \times \Delta y = 
     2.5\sigma\times2.5\sigma$ patches on the lower
		wall (with $u_w=3\varepsilon$); the contrast in $u_w$ 
intrudes into the wall up to a depth of $3\,\sigma$.
	\item a combined topographical and chemical defect (hereafter called ``post''):
       $\Delta x \times \Delta y \times \Delta z =  
			 2.5\sigma\times2.5\sigma\times2.5\sigma$
			 cube protruding from the
		bottom planar wall with an effectively lyophobic chemistry (see the
\emph{Supporting Information} for a precise characterization of this
defect).
\end{itemize}
This setup allows the liquid front to advance or to recede over the defect, while
recovering the undisturbed wedge shape far from it.
Actual defects are certainly more complex; however, already the above model
defects have allowed us to identify the various distinct aspects of hysteresis.

Although devised to produce a liquid wedge, the system adopted in our
calculations and shown in Fig.~\ref{fig:system}B is in the first
instance a narrow channel confined by walls having suitably different
wall-fluid interactions.  In this sense, it is possible to conceive an
equivalent experiment in which a liquid is actually confined between two
chemically distinct solid walls.
While working in the grand canonical ensemble is
appropriate for DFT calculation, the chemical potential can be easily
converted into experimentally relevant observables via
$(\rho_l-\rho_v)(\mu-\mu_0(T))= p_l -p_v$, where $p_l$ and $p_v$
are the liquid and the vapor pressure, respectively.

\subsection{String method}

The string method is used in the context of rare events in order to determine the
path of maximum probability connecting two metastable states \cite{e2002}. While
on its own DFT is capable of capturing \emph{only} the stable and metastable states
by minimizing the relevant free energy (Eq.~\eqref{eq:omega}), the string
method allows one to access also the intermediate configurations of the system along
the activated process -- in this case a liquid wedge advancing or receding
across a surface defect.
The probabilistic meaning of the
\emph{transition path} identified by the string method is that of the
most probable path followed by the activated process under investigation
at fixed thermodynamic conditions ($\mu$ and $T$)
\cite{e2002}. Here the space in which the string is computed is 
the same (discretized) density field $\rho(\bm r)$ as used in
the DFT calculations. Thus by combining DFT calculations and
the string method we identify the most probable path for an
advancing or receding liquid wedge in terms of the natural descriptor of a
capillary system, \emph{i.e}, the local number density.

The string  $\rho(\bm r, \tau)$ can be thought of as an abstract curve 
such that  for each real number $\tau_0 \in [\alpha,\beta]$ there is a
corresponding 3d number density distribution $\rho(\bm r,
\tau=\tau_0)$.
The curve connects the initial and final states A and B, respectively, of the activated process
of advance and retreat, which are defined as $\rho(\bm r,
\tau=\alpha)$ and	$\rho(\bm r, 	\tau=\beta)$.
In the case
of a liquid wedge, the states A and B physically correspond to the liquid wedge far from the
defect on the left and right, respectively; being (metastable) minima of the free
energy, these states can be computed
by standard DFT calculations (see  Fig.~\ref{fig:system}). The
intermediate configurations forming the transition path are found, instead,
by imposing the constrained free energy minimization
\begin{equation}
	\left(\frac{\delta \Omega[\rho]}{\delta \rho (\bm r)}
	\right)_\perp \;\;\bigg\rvert_{\rho=\rho(\textbf{r}, \tau)} = 0
	\text{ ,}
	\label{eq:MFEP}
\end{equation}
for all points $\rho(\bm r, \tau)$ along the string \cite{e2002}.
Equation~\eqref{eq:MFEP} states that the components of the functional
derivative of the grand potential with respect to the density
perpendicular to the string (denoted as $\perp$) have to be zero. In the
direction tangential ($\parallel$) to the string the driving force $(\delta
\Omega[\rho]/\delta \rho(\bm r))_\parallel$ can be
non-zero\footnote{This driving force arises from the fact that, away
from A and B, the system
is not at equilibrium; therefore this force is of ``intrinsic'' character and 
thus should be distinguished from the \emph{external} driving force which will
be discussed in the next sections.}. In short, 
condition~\eqref{eq:MFEP} requires the transition path to lay at the
bottom of a ``valley'' in the free energy landscape, so that the only
intrinsic driving force for the process is the one acting along that valley.  
The initial and final states A and B, instead, are (local) minima of the free energy landscape where
all components of the functional derivative vanish, $\delta
\Omega[\rho]/\delta \rho(\bm r) =0$, and there is no intrinsic driving force.

The string is discretized in terms
of so-called \emph{images}, which \emph{de facto} are number density
configurations at distinct stages of the advancement of the wetting
process.  The initial and final images, corresponding to the
(meta)stable states A and B, are kept fixed.  We employ the improved
string method due to E, Ren, and Vanden-Eijnden \cite{e2007}, which converges to Eq.~\eqref{eq:MFEP} upon
iterating the following algorithm:
\begin{enumerate}
	\item Evolution of the images for a few minimization steps ($5$ to
		$20$) of the grand potential~\eqref{eq:omega}. Here we use the standard Picard
		iteration technique frequently used within classical DFT \cite{roth2010}.
	\item Reparametrization of the string in order to enforce an equal, generalized 
		``distance'' (in the abstract function space)
               {$d= \sqrt{\int\; \mathrm d^3 r \;(\rho(\bm r,
		\tau_i) -\rho(\bm r, \tau_{i+1}))^2}$} between contiguous images 
                numbered by $i$ (corresponding to an
		equal arc length parametrization).
\end{enumerate}
The first step, amounting to a standard minimization procedure, ensures that
the images evolve towards the bottom of the valleys of the free energy
landscape. The second step is implemented via the method of Lagrange
multipliers which enforces a uniform parametrization of the images along the
string by constraining $d$. In other words, the second step prevents
the images to converge towards the closest minimum (states A or B) as it would
happen in a standard DFT calculation consisting solely of the first step.  Further details of the
combined DFT and string method are given in the \emph{Supporting Information}.

%%%%%%%%%%%%%%%%%%%%%%%%%%%%%%%%%%%%%%%%%%%%%%%%%%%%%%%%%%%%%%%%%%%%%

%%%%%%%%%%%%%%%%%%%%%%%%%%%%%%%%%%%%%%%%%%%%%%%%%%%%%%%%%%%%%%%%%%%%%

\section{Liquid wedge migrating across a nanodefect}

%%%%%%%%%%%%%%%%%%%%%%%%%%%%%%%%%%%%%%%%%%%%%%%%%%%%%%%%%%%%%%%%%%%%%

\subsection{Transition path}

\begin{figure*}[tpb!]
	\centering
	\includegraphics[width=\textwidth]{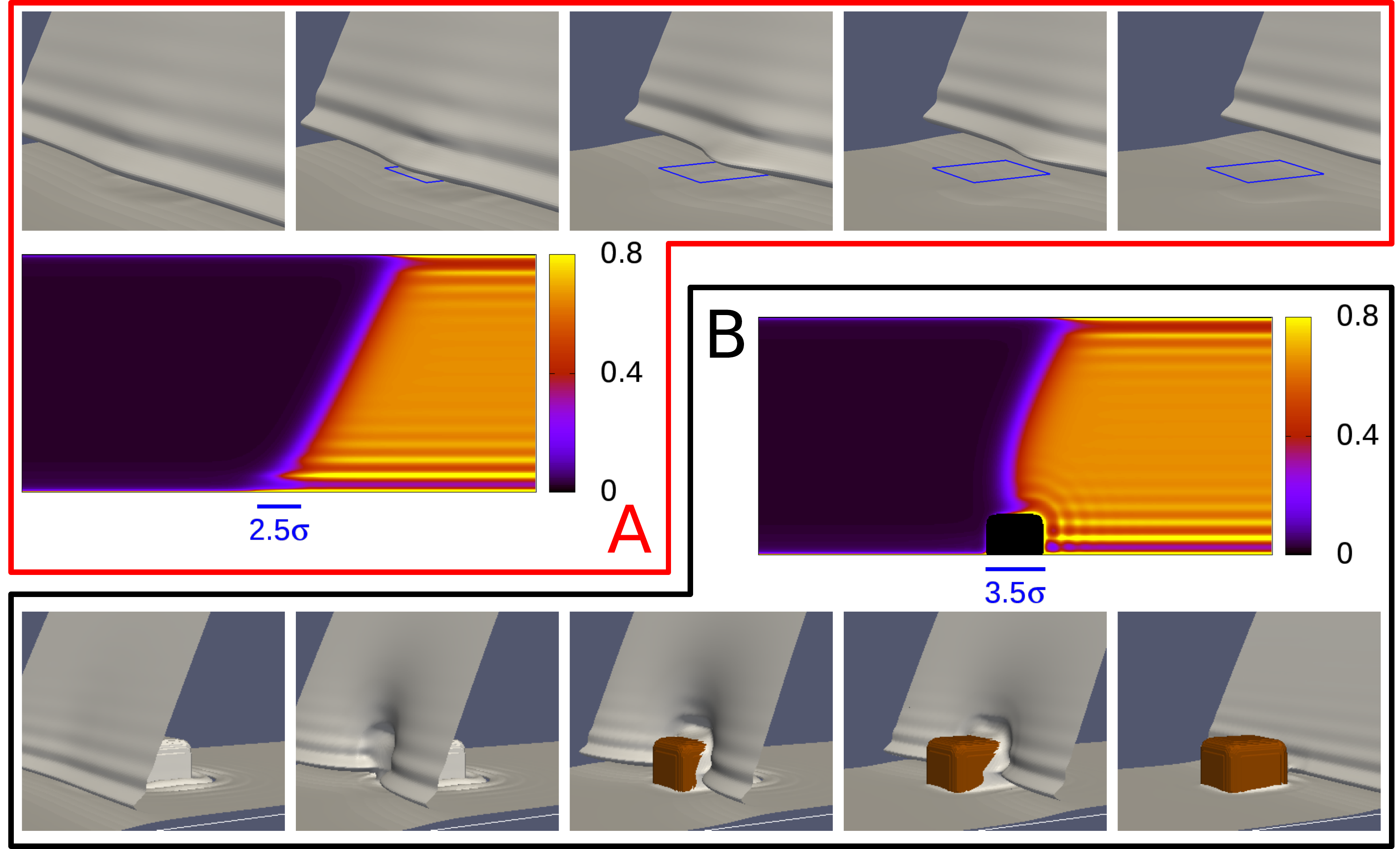}
	\caption{Transition path at bulk coexistence $\mu =\mu_0(T)$
		for both the advance and the retreat across the completely wet patch with $u_w=4\varepsilon$  (A)
		and the cubic post (B). In the top and bottom rows, the gray
		iso-density surface $\rho^\ast$ locate the
		liquid-vapor interface; the liquid domain lies on the right of these
		surfaces and the vapor on the left. In the top row the
		position of the wet patch is indicated by the blue
		square.
		In the bottom row, the brown color corresponds to the post,
		uncovered by liquid.
		In the middle row the number density $\rho(\bm r) \sigma^3$ is shown
		for a single image of the string corresponding to the third picture
		both for the chemical patch (top row) and for the post defect (bottom row); 
		the section is taken in the center of the channel.
		The bars in the middle row denote the lateral position and the size of the defects.
	\label{fig:path}}
\end{figure*}

In Fig.~\ref{fig:path} we show the transition
paths for the completely wet chemical defect and for the post
defect. In order to visualize the three-dimensional deformation
of the interface as it crosses a defect we use the iso-density
surface $\rho^\ast=(\rho_l+\rho_v)/2$, where $\rho_l$ and $\rho_v$
denote the bulk liquid and vapor density, respectively; 
this surface (in gray) reveals that upon retreat the lower wall is
covered by the expected liquid-like wetting film \cite{getta1998,bauer1999}.
The post, instead, dries upon retreat (in brown), behaving as an
effectively lyophobic chemical defect.

%Fig2

Based on macroscopic arguments it is expected that the
interface deformations decay exponentially away from the defect 
\cite{joanny1984}. Figure~\ref{fig:path} shows that the decay is
indeed rapid even on the nanoscale and thus
explains the difficulty to experimentally resolve the deformations
induced by a single nanodefect. This result also suggests that,
although the periodic boundary conditions in principle mimic an
array of defects, the effect of having a periodic arrangement is limited and 
does not hinder the global advancement or retreat of the liquid-vapor
interface.

The two defects have a qualitatively different effect on the liquid
wedge: in the case of the wet patch, the receding liquid domain is held
back by the defect, while in the post case it is locally pushed towards further
retreat. On the other hand, the advancing liquid domain is pulled
forward by the wet patch and is held back by the post.  Accordingly, the
deformations of the liquid-vapor interface along the
transition paths suggest that the completely wet patch (post) hinders
the receding (advancing) motion of a liquid front. 
This will be clarified in terms of free energy barriers in the next
subsection.

It is important to note that the transition paths shown in
Fig.~\ref{fig:path} do \emph{not} have a directionality: they are the
same for advancing and receding liquid wedges at fixed
external conditions $\mu$ and $T$. As will be shown in the
following, hysteresis arises because along a given transition path the system might be trapped in 
a metastable minimum. Whether this minimum 
occurs or not depends on the forces driving the system; \emph{e.g.}, tuning
the chemical potential $\mu$ off coexistence drives the liquid wedge externally. 

A representative density field is shown in the middle row of
Fig.~\ref{fig:path} for configurations for which the liquid front is just
about to cross the defect. The liquid layering induced by the walls is
apparent, with more pronounced effects at the lower lyophilic wall.
The patch, which is more lyophilic than the rest of that wall,
further enhances the density oscillations at the wall and shifts their
maxima towards the wall (see, \emph{e.g.}, Ref.~\cite{nold2014}). Close to the
post, instead, the layering is slightly reduced due to its lyophobic
nature, which is reflected by the local contact angle $\theta_Y\approx 125^\circ$ on the top of the
defect.
At the lower corner of the cube the interference of packing
effects on the horizontal surface and on the sidewalls induces regions
where the density is alternatingly increased and reduced.

\subsection{Free energy profiles}

%Fig3
\begin{figure}[t!]
\centering
\includegraphics[width=0.6\textwidth]{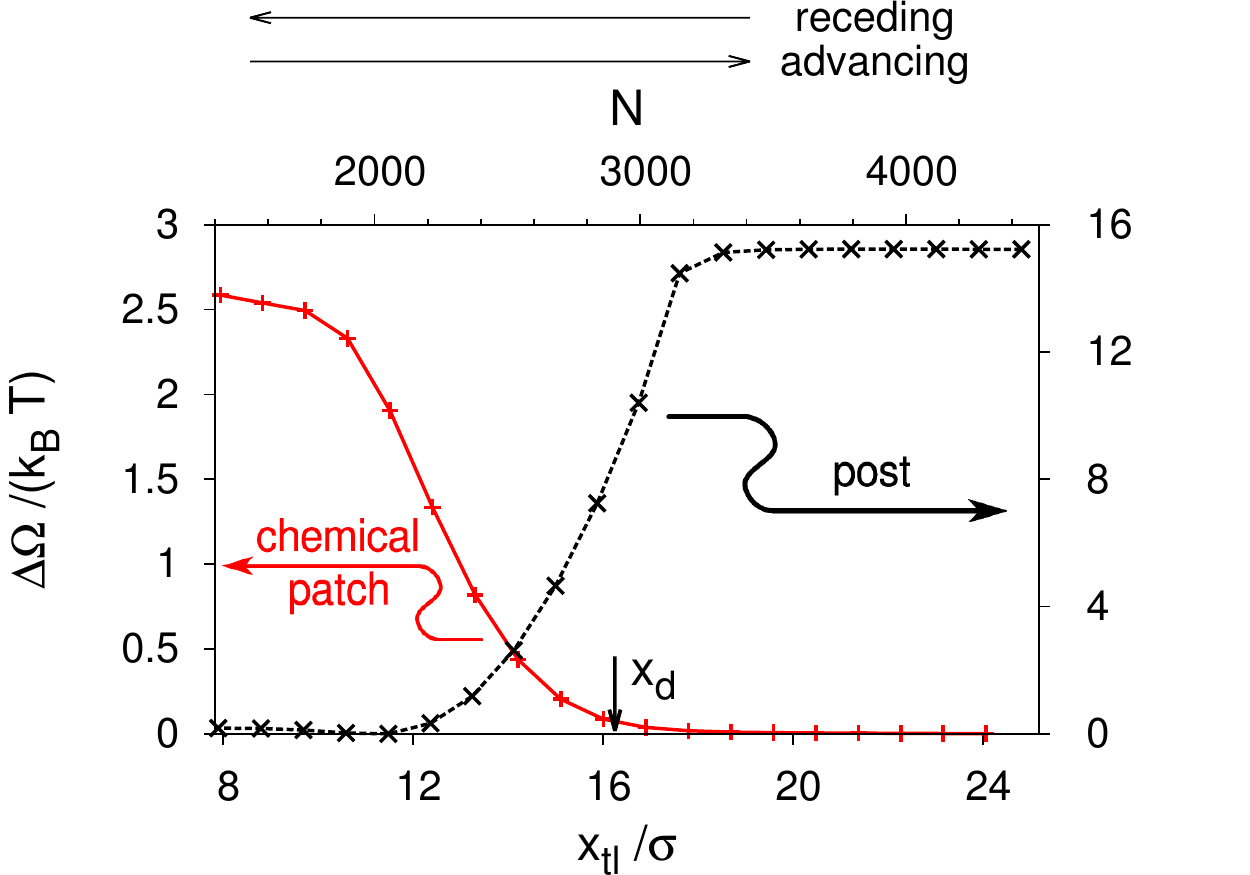}
\caption{Excess free energy profiles $\Delta \Omega(x_{tl})$ at two-phase coexistence
	($\mu=\mu_0(T)$) computed from the transition paths shown in
	Fig.~\ref{fig:path};
	$x_{tl}$ is a measure of the position of the triple line defined
	as $x_{tl}= N /(L_y L_z (\rho_l -\rho_v))$, where $N$ is the number of
	fluid particles in the computational box.
	Receding (advancing) corresponds to a decrease (an increase) of
	$x_{tl}$ or $N$ (see Fig.~\ref{fig:system}B).
	The defects considered here are a completely wet chemical  
	patch (red solid line,
	$u_w=4\varepsilon$) and a cubic post (black dashed line)
  placed at the horizontal position $x_{d} = 16.25\sigma$.
	After subtracting the defect-free free energy profile, in both cases the resulting free energy 
	difference $\Delta \Omega(x_{tl})$ exhibits an upper and a lower plateau.
	$\Delta \Omega(x_{tl})$ is shifted such that the latter one is zero. 
	For the wet chemical patch, the thermodynamically stable state (absolute
	minimum) is outside of the considered range of $x_{tl}$. For the post, a
	very shallow minimum is attained at $x_{tl}\approx 11$; since, within our
	computational precision, this minimum cannot be distinguished from the
	plateau, we regard the lower part of the post profile as a plateau
  with $\Delta \Omega = 0$. 
	\label{fig:profiles}}
\end{figure}

From the density fields constituting the transition path
(Fig.~\ref{fig:path}) and from Eq.~\eqref{eq:omega} we compute the free
energy profiles  for the advance or retreat of the liquid wedge along
the string. 
In Fig.~S4 we demonstrate that there is a one-to-one
correspondence between the image number along the string and the amount of fluid in the
channel, \emph{i.e.}, the number of particles $N=\int\;\mathrm{d}^3 r \; \rho(\bm
r)$ as obtained along the transition path. 
An even more instructive parametrization of the string is the quantity 
$x_{tl}= N /(L_y L_z (\rho_l -\rho_v))$, where $L_y$ and $L_z$ are the dimensions of the computational box in the
$y$ and $z$ direction, respectively; it serves as a measure of the position of
the triple line.
 
From the free energy profiles for the advance or retreat of the liquid wedge
over a defect we subtract the free energy profile obtained for the filling
of the corresponding defect-free channel (Figs.~\ref{fig:system}B and
S5).  This results in the excess free energy profile $\Delta
\Omega(x_{tl})$ for
a system with zero external driving forces which is plotted in
Fig.~\ref{fig:profiles}.  The data show a monotonic trend of the free
energy profile connecting the stable and metastable states corresponding
to the lower and upper plateau branches, respectively. The occurrence of these
essentially horizontal plateaus shows that the interaction of the liquid
wedge with a defect decays rapidly upon increasing the distance between them.  

In the case of a wet (chemical) patch, the stable state corresponds to a configuration
in which the liquid completely covers the defect and the liquid-vapor
interface is undeformed and far to the left
of the patch ($x_{tl}$ large); the
metastable state corresponds to the patch being exposed to the vapor and
the planar liquid-vapor interface far to the right of the patch ($x_{tl}$ small).
The intermediate configurations are characterized by a progressive
increase of the free energy, due to the wetting of the defect and the
deformation of the liquid-vapor interface. Thus the wet patch
constitutes an obstacle to the retreat of the liquid wedge,
characterized by a free energy barrier of $2.5$~$k_BT$
(Table~\ref{tab:barriers}). Viewing the free energy profile in the
opposite direction, the wet patch favors the advancing motion of the liquid
front.

The post represents, instead, a hindrance to the advance of the liquid
wedge with a free energy barrier of $15$~$k_BT$. Since according to
transition state theory the typical time 
$t_{cr}$ for \emph{cr}ossing a free energy
barrier $\Omega_{barrier}>0$ scales exponentially with its height,
$t_{cr}=t_0\exp{(\Omega_{barrier}/(k_B T))}$, a single post may block
the advance of a liquid front on a time scale $t_{cr}\approx 0.5$~$\mu$s
which is relevant for experimental observations (at least at
$\mu=\mu_0(T)$; for $\mu$ being off coexistence see below). For this estimate we 
have adopted conservatively a
microscopic attempt frequency $1/t_0=k_B T/h\approx
10^{13}$~s$^{-1}$, where $h$ is Planck's constant \cite{blake1969}.

Given the sigmoidal shape of the free energy profiles in
Figs.~\ref{fig:profiles} and S5, the quantity characterizing thermally
activated (de)pinning by (from) various types of defects is the free
energy difference $\delta \Omega \equiv \Omega_{wet}-\Omega_{dry}$ between
the two plateau values, where 
$\Omega_{wet}$ and $\Omega_{dry}$ are the free energy of the state in which
the liquid-vapor interface is sufficiently far from the defect to the
left and to the right (Fig.~\ref{fig:system}), respectively. 
The free energy differences $\delta \Omega$ are reported in
Table~\ref{tab:barriers}. 
A positive sign of $\delta \Omega$ implies that an advancing liquid
front has to climb up a barrier and is hindered in its motion
whereas for $\delta \Omega$ negative a receding liquid front
has to climb up a barrier and is hindered. 
The data in Table~\ref{tab:barriers} tell that,
if at all, single nanometer-sized chemical blemishes are only weak
obstacles to the motion of a liquid front: thermal fluctuations can be sufficient to
facilitate the crossing of a single chemical defect. 
On the other hand, nanodefects which are both topographical and
chemical such as the post strongly hinder the
motion of liquid fronts.

The ratio between the free energy barrier
$\Omega_{barrier}\equiv\lvert \delta \Omega \rvert$ and the thermal
energy $k_B T$ discriminates between deterministic pinning and reversible
thermally activated motion of the triple line:
for $\Omega_{barrier}/(k_B T)\sim 1$ the defect is weak and thermal
fluctuations are able to \emph{reversibly} switch between the two
plateaus, while for $\Omega_{barrier}/(k_B T)\gg 1$ the liquid front is
blocked by the strong defect.
This measure of the defect strength is an alternative to the classical distinction of weak and
strong defects based on the ``spring constant'' and the wetting force
of the defects \cite{joanny1984}. 
Our findings seem to be in accordance with the experiments of Delmas
\emph{et al.} \cite{delmas2011}, who find that the crossing of
nanodefects is reversible if the defect ``energy'' is comparable with
$k_B T$.
For high barriers, external forces have to be applied in order to drive the
liquid front across the defects. As discussed in the next section, this
requires knowledge of the full free energy profile. The same is true
for intermediate barriers in which case force assisted, thermally
activated (de)pinning may become relevant.

%Table1
\begin{table}[tb!]
	\centering
	\caption{Summary of the results for the advance and retreat of a
	liquid wedge across defects: wall energy $u_w$ for the \emph{up}per and \emph{low}er walls
	and the defects, Young contact angle
	$\theta_Y$, free energy difference at two-phase coexistence $\delta
	\Omega \equiv \Omega_{wet}-\Omega_{dry}$, corresponding macroscopic estimates $\delta
	\Omega^{\mathrm{macro}}$, excess chemical potential at the spinodal $\Delta \mu_{sp}$,
	and excess free energy at the spinodal
	$\Delta\Omega_0(x_{tl,sp})\equiv \Delta\Omega(\Delta
	\mu=\mu-\mu_0(T)=0,V,T;x_{tl,sp})$  
	(see next sections). A positive sign of $\delta
	\Omega$ implies a barrier $\Omega_{barrier}=\delta \Omega$ for an
	advancing liquid front, a negative sign implies a barrier
	$\Omega_{barrier}=|\delta \Omega|$ for a receding liquid front.
	$\theta_Y$ is determined independently via the interfacial tensions using
	Young's formula
	$\cos\theta_Y=(\gamma_{sv}-\gamma_{sl})/\gamma_{\,lv}$; because of an imperfect choice of the wall potentials,
 	$\theta_Y^{\mathrm{up}}+\theta_Y^{\mathrm{low}}=180.18^\circ > 180^\circ$.
	}	
\label{tab:barriers}
\vspace{0.5cm}
\begin{tabular}{lcccccc}
	defect type&$u_w/\varepsilon$&$\theta_Y$ &
	$\delta \Omega$& $\hspace{0.7cm}{\delta \Omega^{\mathrm{macro}}}$ &
	$\underline{\Delta \mu_{sp}}$ & $\underline{\Delta\Omega_0(x_{tl,sp})}$\\
	& & &$\overline{k_B T}$ &$\overline{k_B T}$& $k_B T$ &$k_B T$ \\
\hline
defect-free channel (up)& $2.6$ & $99.67^\circ$ 	& --- 	& --- 	& --- & --- \\
defect-free channel (low)& $3.0$ & $80.51^\circ$ 	& --- 	& --- 	& --- & --- \\
lyophobic patch& $2.0$ & $124.85^\circ$ 	& $2.4$  & $1.87$ 		&$0.0035$ 		& $1.1 $ \\
partially wet patch& $3.5$ & $51.13^\circ$ 	& $-1.3$ & $-1.17$  & $-0.00186$  & $0.61$ \\
completely wet patch& $4.0$ & $0.0^\circ$ 	& $-2.5$ & $-2.22$  & $-0.0034$  	& $1.2 $ \\
cubic post& $3.0$ & $126.81^\circ$& $15.2$ & $8.02$  & $0.027$  		& $10.5$   \\
\hline
\end{tabular}
\end{table}

Within a macroscopic theory the free energy differences $\delta \Omega \equiv \Omega_{wet}-\Omega_{dry}$ 
can be easily estimated \emph{a priori} from the surface free energy associated with
the defect: for macroscopic chemical blemishes, for which the
relevant quantities (the solid-liquid and the solid-vapor interfacial tensions) are well defined and line
contributions do not matter, the free energy difference is
$\delta \Omega^{\mathrm{macro}} = \gamma_{\,lv}
\left(\cos{\theta_Y^{\mathrm{sub}}} - \cos{\theta_Y^{\mathrm{def}}}\right) A$,
where 
$\theta_Y^{\mathrm{sub}}$ and $\theta_Y^{\mathrm{def}}$ is 
Young contact angle related to the defect-free \emph{sub}strate and to the
one on the \emph{def}ect, respectively, and $A$ is the area of the
defect. In the case of a topographical defect one has 
$\delta \Omega^{\mathrm{macro}} = -\gamma_{\,lv}\cos{\theta_Y} A_{lat}$, 
where $A_{lat}$ is the area of the side walls of the
protrusion; $\theta_Y$ is Young's angle which, for a purely topographical defect 
and on a macroscopic level, is the same for the substrate and
the defect. 
For a topographical and chemical defect, such as the post, the
macroscopic expression is a combination of the previous two: 
$\delta \Omega^{\mathrm{macro}} = \gamma_{\,lv}
\left(\cos{\theta_Y^{\mathrm{sub}}} - \cos{\theta_Y^{\mathrm{def}}}\right) A
-\gamma_{\,lv}\cos{\theta_Y^{\mathrm{def}}}A_{lat}$.
The macroscopic estimates
$\delta \Omega^{\mathrm{macro}}$ are reported in
Table~\ref{tab:barriers}. For the chemical patches these agree surprisingly well 
with the results from the corresponding microscopic calculations. For the
post defect, however, the macroscopic estimate is about one half of that obtained
from the microscopic theory. 
One reason for the inaccuracy of the macroscopic estimate is the particular liquid
structure near the nanosized post, which is not captured by a macroscopic
description. The other reason seems to be related to  the weaker attraction of the
walls of nanosized posts in comparison with the walls of macroscopic bodies
composed of the same material.

\subsection{Forcing the system: the origin of hysteresis}

Free energy profiles at two-phase coexistence \emph{per se} do not explain the
emergence of wetting hysteresis. Hysteresis is due to the occurrence of
a free energy barrier between a metastable ``pinned'' state and the thermodynamically stable
state. The ``pinned'' state is present only within a certain range of 
external forcing, here realized via the chemical potential $\mu$.
Experimentally, the same forcing can be obtained 
by small temperature differences between the sample
and the liquid reservoir, which controls $\Delta \mu$ for volatile
liquids \cite{hofmann2010}, or by changing the pressure (for non-volatile liquids).
Body forces like gravity, electrical fields, etc. act in a similar way.
Depending on the history of the forcing -- \emph{i.e.}, the actual
experimental procedure -- the system can either get trapped or not in
metastable states and thus can evolve through different sequences of
configurations while advancing or receding.

% Fig4   
\begin{figure}[t!]
\centering
\includegraphics[width=0.7\textwidth]{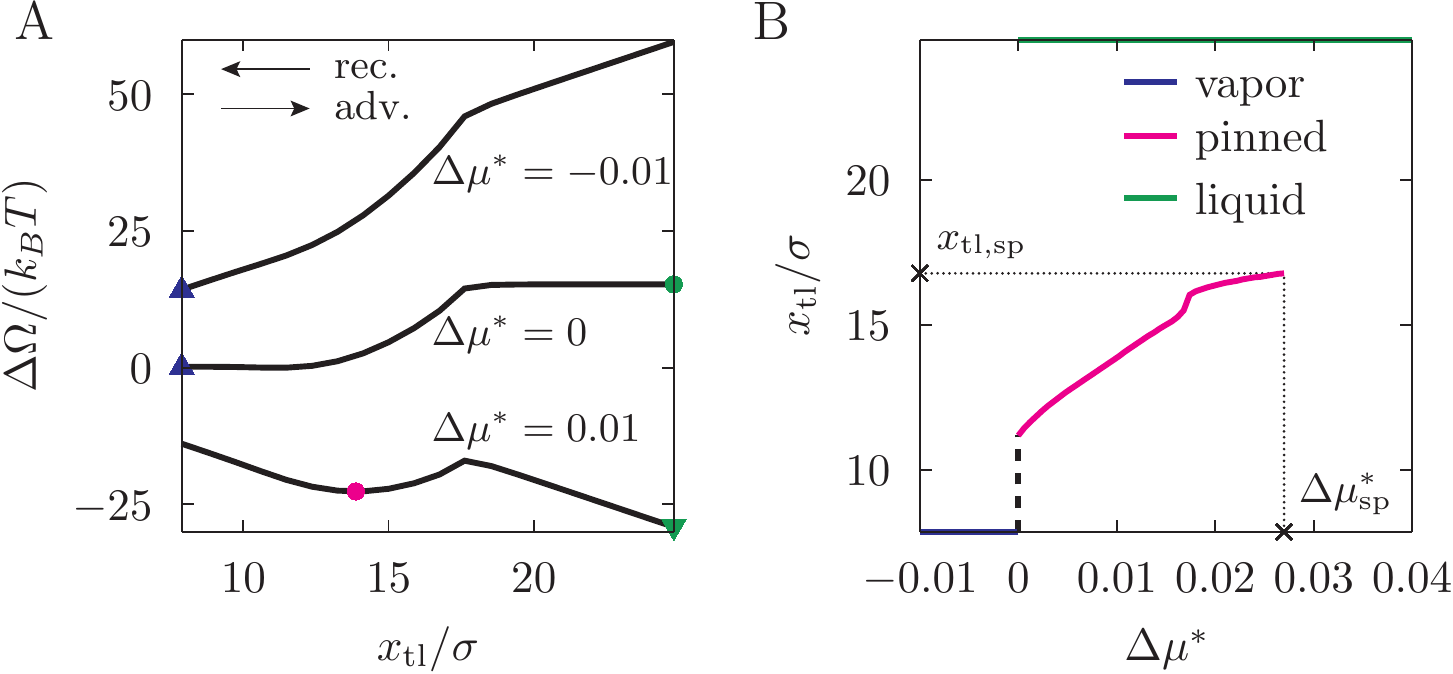}
\caption{\emph{A}) Excess free energy profiles $\Delta \Omega$ for 
	selected reduced chemical potentials $\Delta \mu^\ast=(\mu-\mu_0(T))/
	\varepsilon$ for the cubic post as a function of the
	position $x_{tl}$ of the triple line. 
	Symbols identify the local (circle) and global minima (triangles) for a given $\mu$.
	The global minima are outside of the plotted range at $x_{tl,v}$ and
	$x_{tl,l}$ (see main text); 
	here the corresponding symbols are drawn at the edge of the
	computational box.  
	\emph{B}) 
	Position $x_{tl}$ of the triple line at the minima of the free
	energy profiles in \emph{A} as a function of $\Delta\mu^\ast$: metastable ``pinned'' state
	(magenta), stable vapor phase (blue), and stable liquid phase (green).
	The pinned minimum disappears at the spinodal $\Delta \mu_{sp}^\ast$.
	The black dashed line indicates that for $\Delta \mu^\ast = 0^+$ the pinned
	minimum is likely to shift continuously towards the vapor phase. However, the very steep slope
of this black line prevents to	verify this behavior in the actual
calculations.
	\label{fig:forced}}
\end{figure}

In order to reveal this mechanism for hysteresis, for the defects
introduced above we compute the transition path and the free energy
profiles for several values  of $\mu$ off coexistence, as done in
Fig.~\ref{fig:profiles} for $\mu=\mu_0(t)$. 
For the post defect, Fig.~\ref{fig:forced}A shows the excess free energy profiles as a function of
the position $x_{tl}$ of the triple line. As expected,
positive reduced chemical potentials $\Delta \mu^\ast\equiv (\mu
-\mu_0(T))/\varepsilon$ tilt the whole free energy profile towards
the fully wet state,  so that at $x_{tl}=x_{tl,l}(\mu)$ the liquid phase fills the whole space of the channel
(green triangle).\footnote{In the case of an unbounded liquid wedge, one
has $x_{tl,l}\to \infty$. However, for the computational box shown in
Fig.~\ref{fig:system}, $x_{tl,l}$ is finite and corresponds to the
position of the triple line at the front end of the computational box.}
For $\Delta \mu^\ast>0$ the sigmoidal free energy profile observed for $\Delta \mu^\ast=0$
(the same as the dashed black line in Fig.~\ref{fig:profiles}) turns into one featuring a 
metastable state (magenta circle, emerging from the blue
triangle), separated from the stable one (green triangle) 
by a barrier. 
The metastable state (magenta circle) corresponds to one with the meniscus in the channel pinned at
the defect; it exhibits a deformation of the liquid-vapor interface and a
perturbation of the density distribution as compared with the defect-free configuration (see Fig.~\ref{fig:path}).
This metastable state occurs within the range $0<\Delta \mu^\ast<\Delta
\mu^\ast_{sp}$, with $\Delta \mu^\ast_{sp}$ as the spinodal value. 
In this range the thermodynamically stable equilibrium (i.e., the global
minimum of the free energy profile) corresponds to the liquid phase
(green triangle); 
however, free energy barriers of ca. $10\,k_BT$ may effectively trap the system in the 
metastable ``pinned'' state, thus giving rise to hysteresis (see
Ref.~\cite{checco2014} for a similar situation on superhydrophobic surfaces).
For $\Delta \mu^\ast<0$, instead, the only stable state is the one 
filling the computational box with vapor (blue triangle).
In Fig.~\ref{fig:forced}B we show the loci of the free energy minima
(stable or metastable ones) in the $\Delta \mu^\ast$-$x_{tl}$ plane.

The calculations off coexistence also show that the transition paths
(\emph{e.g.}, the sequence of number densities shown in
Fig.~\ref{fig:path}) are not changed by the external ``force''
$\mu$ 
(for an extended discussion, see Refs.~\cite{Giacomello2012,giacomello2015}).
Thus, given the weak dependence of
$\rho(\bm r, \tau)$ on $\mu$ for fixed $x_{tl}$,
it follows from Eq.~\eqref{eq:omega} that the free energy profiles
(i.e., the sequence of constrained free energies along
the most probable path, which is parametrized, e.g., by the
position $x_{tl}$ of the triple line)
$\Omega(\mu,V,T;x_{tl})$ off coexistence
can be computed within good approximation from that at coexistence:  
\begin{equation}
	\Omega(\mu,V,T;x_{tl})=\Omega(\mu_0(T),V,T;x_{tl}) - (\mu-\mu_0(T))
	N(x_{tl}) \text{ ,}
	\label{eq:shift}
\end{equation}
where the relation between the number of particles and the
position of the triple line is $N(x_{tl})=x_{tl} L_y L_z (\rho_l -\rho_v)$.
Equation~\eqref{eq:shift} states that the only relevant effect of tuning the
chemical potential off coexistence is to tilt the free energy profile
obtained at $\Delta \mu^\ast=0$. In that way metastable minima separated
by barriers from the stable ones may be created along the transition
paths. In Fig.~S6, it is demonstrated that free energy profiles
off coexistence obtained via a direct calculation or from
Eq.~\eqref{eq:shift} actually coincide.

%Fig5
\begin{figure}[t!]
\centering
\includegraphics[width=0.6\textwidth]{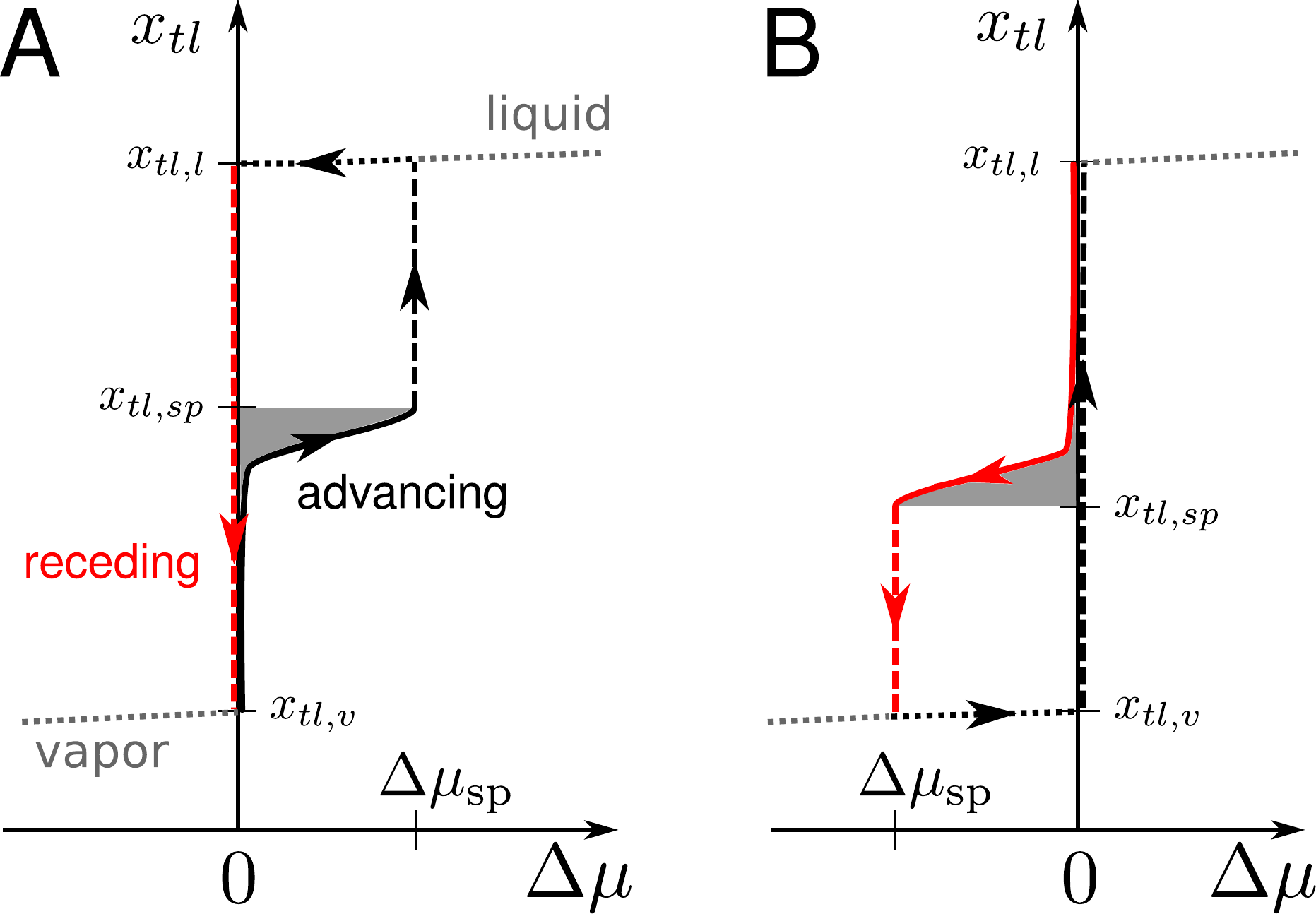}
\caption{Sketch of the behavior of an advancing or receding liquid wedge
across defects of the post type (\emph{A}, compare with
Figs.~\ref{fig:profiles} and \ref{fig:forced}B: advance barrier, spinodal $\Delta
\mu_{sp}>0$) and of the wet chemical patch type (\emph{B}, retreat barrier
Fig.~\ref{fig:profiles},
$\Delta \mu_{sp}<0$). The sequence of (meta)stable states
constituting the advancing (receding) process is plotted with a solid
\emph{black} (solid \emph{red}) line. The loci of the minima of the free energy
related to the pure liquid or pure vapor phases are shown as dotted black
lines. Dashed lines are used for irreversible processes, \emph{i.e.},
the ``snapping'' of the contact line at $\Delta \mu_{sp}$ and the
sliding of the liquid wedge downhill the free energy profiles at $\Delta \mu
=0$. 
The area shaded in gray,
$\int_{\mathit{x}_{tl,v}}^{x_{\mathrm{tl,sp}}} \; \Delta \mu \; \mathrm d x_{tl} $
	in \emph{A},
	$\int_{\mathit{x}_{tl,l}}^{x_{\mathrm{tl,sp}}} \; \Delta \mu \; \mathrm d	x_{tl} $
	 in \emph{B},
	 corresponds to  $\Delta \Omega_0(x_{tl,sp})/(L_y L_z (\rho_l -\rho_v))$; this can be obtained using  
         Eq.~\eqref{eq:CAHrow-a} in the above expressions
				for the areas. 
\label{fig:hysteresis}}
\end{figure}

The overall mechanism of hysteresis during the advance or retreat of a
liquid wedge is summarized in Fig.~\ref{fig:hysteresis}. To this end we
consider an experimental procedure according to which the system is initially in
the vapor phase (lower left corner of Fig.~\ref{fig:hysteresis}A) and
the chemical potential -- or, alternatively, the pressure --
is gradually increased (black line); the
experiment is assumed to be carried out adiabatically in order to allow
the system to relax to the local minimum of the free energy.
At coexistence $\Delta \mu=0$ a
liquid-vapor interface is formed, such that it is assumed to be formed 
on the right of the post defect (see Fig.~\ref{fig:path}B) and
moves to the left along the wall up to the post, where it is stopped by the free
energy barrier. Upon increasing $\mu$ the interface remains pinned at
the defect until the free energy barrier has decreased to a value
comparable with $k_B T$, which occurs near the spinodal
$\mu=\mu_{sp}$. 
For $\mu>\mu_{sp}$ the channel is completely filled by liquid.  The retreat
of the wedge (red line) follows a different sequence of states, because
the system remains trapped in the global minimum (liquid state). The
liquid-vapor interface is formed only at coexistence $\Delta \mu=0$, at
which the wedge moves unhindered along the wall to the other minimum,
\emph{i.e.}, the vapor state. The two qualitatively different advancing
and receding branches in Fig.~\ref{fig:hysteresis} explain wetting
hysteresis induced by surface defects. 

The range of chemical potentials $\Delta \mu$, for which the metastable pinned state
occurs, is for the post case $0\leq \Delta \mu^\ast \leq \Delta \mu_{sp}^\ast=0.027$ in units of
$\varepsilon$ (for the other defects see Table~\ref{tab:barriers}).
However, $\Delta \mu_{sp}^\ast$ is not an intrinsic quantity as it depends on the individual system under
consideration and, in particular, on the box dimensions.
For a given value of $\Delta \mu$ the total force exerted on the liquid-vapor 
interface increases proportionally to the cross-section of the channel.
This force has to be balanced by the single defect in the box.
However, the maximum force the individual defect can withstand is given
and independent of the computational box size. Thus, upon doubling the height
of the computational box, $\Delta \mu_{sp}^\ast$ is halved.   
Therefore, in the following section defect-specific quantities independent of incidental
computational details are introduced which allow us to relate the characteristics
of a specific defect to its macroscopic effect on contact angles.

We are now in a position to compare in more detail the present approach with the
seminal paper of Joanny and {de Gennes} (JdG) \cite{joanny1984}. This
comparison is beneficial in order to understand the differences between wetting
hysteresis on the macroscale and on the nanoscale.
First, JdG consider a special case of wetting hysteresis,
i.e., one which arises from macroscale ($>30$~nm), \emph{strong} defects,
where strong means that the defect is capable of inducing two distinct 
configurations of the liquid-vapor interface -- a weakly and a strongly 
deformed one -- at the 
same nominal position of the unperturbed contact line. Depending on the history of
contact line motion (i.e., advance or retreat) one of these two configurations
is selected, giving rise to wetting hysteresis. However the nanodefects 
considered here are weak in the JdG sense,
and there is only a single configuration of the liquid-vapor interface
for a given position of the triple line, as demonstrated by the
free energy profiles in Figs.~\ref{fig:profiles} and
\ref{fig:forced}A.
In fact, wetting hysteresis for these nanodefects emerges if the quasi-static 
procedure in Fig.~\ref{fig:hysteresis}
requires different forcing for advance and retreat. Alternatively,
it can be regarded as a hysteresis in the contact line position upon cycling
the driving force.
In our approach we explicitly refer to the driving force $\Delta \mu$
(or $\Delta p$);
this suggests an experimental realization of our thought experiment in
which a liquid front is quasi-statically driven through a narrow channel
with suitably prepared walls by stepwise changing the pressure in an
attached fluid reservoir and simultaneously measuring the position of
the liquid front. 
Alternatively, the AFM technique with special carbon tips used in Ref.~\cite{delmas2011} is capable
to determine both the force and the position of the contact line for a
single nanodefect; a comparison with Fig.~\ref{fig:hysteresis} appears
to be possible.\footnote{For a direct comparison, an experimental characterization of the 
nanodefect is needed. In addition, the comparison should be done not
directly with $\Delta \mu$, but with the (intrinsic) defect force
$\partial \Omega/\partial x_{tl}$ introduced in the next section.}
Another experimental technique is the noncontact AFM for nanoscale
wetting used by Checco \emph{et al.}~\cite{checco2006}. On a larger scale, also the Wilhelmy plate
technique \cite{priest2007} could be used, while macroscopic contact angle
hysteresis arising from dilute distributions of nanodefects can be measured and related to the
defect characteristics via Eq.~\eqref{eq:CAH} (see next section).
Another peculiarity of nanodefects are thermally activated crossings,
the description of which requires specialized rare-events techniques in order to reveal
the transition path and thus the (unexpected) sigmoidal free energy profile.
In a certain sense, the findings of the present nanoscale analysis are
complementary to the macroscopic picture of JdG for strong defects, revealing that 
wetting hysteresis induced by nanodefects proceeds via a different mechanism;
hysteresis disappears if thermal fluctuations become dominant.

A major difference between the two approches is our capability to
achieve a
microscopic description which can cope with the problem that for
nanodefects there is no clear scale separation between the linear
extensions of the defects and the length scales characterizing the
fluid inhomogeneities at the solid-fluid and at the liquid-vapor interfaces,
the deformations of the latter, and the range of the force fields. 
Indeed we have found that incorporating such details can be important for 
obtaining a quantitative prediction of the pinning characteristics of nanodefects.
Further, in order to fully exploit the potential of our microscopic
approach we are searching for the transition path in the infinite dimensional
configuration space spanned by all possible density distributions.
This way we also avoid the problem of ``hidden'' variables which may arise
if one uses only a reduced set of variables (actually two in the treatment
of JdG); the actual path may be even completely concealed within a reduced
description (see, e.g., Refs. \cite{giacomello2015} and
\cite{bolhuis2002}).  
This latter problem affects the estimates of, e.g., the free energy barriers
involved and, with exponential sensitivity, those of the kinetics.

\section{Advancing and receding contact angles}

In the discussion above, we have referred to \emph{wetting hysteresis}
as the qualitative differences in the advancing and receding processes,
occurring in nanochannels and for a single defect -- thereby avoiding the notion of
contact angle hysteresis. The latter requires scale separation between the
characteristic size of the defects and that of the macroscopic liquid body 
(\emph{e.g.}, the drop), which for nanodefects is typically 
the case, and a definition of the macroscopic apparent 
contact angle $\theta^\mathrm{macro}$. In the following, we provide such a definition and we express
 $\theta^\mathrm{macro}$ in terms of the microscopic quantities given
in Table~\ref{tab:barriers} which, rigorously speaking, were computed for
nanodefects in nanochannels. We first derive these relations for
a straight row of nanoscale defects parallel to the liquid
front. 
Subsequently, we generalize these results to a random, dilute distribution of defects,
which is closer to actual surfaces studied experimentally.

Before discussing the two special arrangements of nanodefects, we 
introduce a general definition of the macroscopic apparent contact angle.
We first observe that the deformations of the liquid-vapor interface
induced by the wall and by the defects vanish at sufficiently large
distances from the walls and from the defects and that the interface  
assumes one of the simple macroscopic shapes compatible with the macroscopic 
Laplace equation (spherical cap, cylindrical cap, plane, etc.). 
This macroscopic interface can be extrapolated up to the wall surface and the 
intersection of these two surfaces
defines the apparent contact angle $\theta^\mathrm{macro}$ and the
macroscopic geometrical position of the three-phase-contact or triple line.  
In the cases discussed here, the macroscopic triple line is a straight line
in the $x$-$y$ plane parallel to the $y$ axis, characterized by the coordinate 
$x = x_{tl}$.   

\subsection{Contact angle hysteresis induced by a row of defects}

In order to characterize \emph{static} contact angle hysteresis, we seek
for mechanically (meta)stable configurations of the contact line.  For
the system studied in the previous sections -- a liquid wedge 
migrating across a periodic arrangement (row) of defects aligned along
the $y$ direction -- metastable states correspond to the minima of
Eq.~\eqref{eq:shift}. 
Differentiating that equation with respect to $x_{tl}$ and recalling 
the relation $\mathrm d N/ \mathrm d x_{tl} = L_y L_z(\rho_l-\rho_v)$
leads to
\begin{equation}
 - \frac{\partial \Omega_0}{\partial x_{tl}}
 \bigg\rvert_{\emph{x}_{tl}=\emph{x}_{tl,st}}
  + L_yL_z(\rho_l-\rho_v)\Delta \mu
  = 0 
\text{ ,}
\label{eq:CAHrow-a}
\end{equation}
where we introduced the more compact notation $\Omega_0
\equiv\Omega(\Delta \mu=0,V,T;x_{tl})$.
The first term is the microscopic force $ - \partial \Omega_0/\partial x_{tl}$ 
exerted on the contact line 
by the single defect present in the computational box. The second term
is the total macroscopic force which has to be balanced by the microscopic
defect force. Using the relation $\Delta \mu (\rho_l-\rho_v) = \Delta p $
the macroscopic force can be also expressed in terms of the pressure difference across the 
liquid-vapor interface as $ L_yL_z \Delta p $. 
As indicated by the acronym \emph{st}, Eq.~\eqref{eq:CAHrow-a}
has \emph{st}able solutions $x_{tl}=x_{tl,st}(\Delta \mu)$
only within a certain range of macroscopic forces along the
advancing or receding processes which correspond to the black or red solid
lines in Fig.~\ref{fig:hysteresis}, respectively. 
The quantity $\partial \Omega_0/\partial
x_{tl}$ is defect-specific and independent of $L_y$ and $L_z$, because
it is determined  by the local wetting properties of the defect and by
the local deformations of the liquid-vapor interface. According to
Eq.~\eqref{eq:CAHrow-a}, in order to render also $\Delta \mu$ and
$\Delta p$ defect-specific, one has to multiply them by
$L_yL_z(\rho_l-\rho_v)$. For instance, by multiplying $\Delta \mu_{sp}$ in 
Table~\ref{tab:barriers} by our system specific quantity
$L_yL_z(\rho_l-\rho_v) \approx 179\sigma^{-1}$
one obtains an intrinsic quantity which can be related to the maximum 
force a defect can exert on the triple line (see Refs. \cite{delmas2011, priest2007}).

In order to connect with macroscopic contact angles and contact angle
hysteresis, we proceed as follows.
First, we observe that in Eq.~\eqref{eq:CAHrow-a} the total macroscopic driving
force $L_yL_z(\rho_l-\rho_v)\Delta \mu =  L_yL_z\Delta p$ must act also on the
contact line. On the other hand, within a macroscopic description, the force on
the triple line is determined by the action of the interfacial tensions. Its relevant lateral
component is Young's unbalanced force $-L_y\gamma_{lv}(\cos\theta^\mathrm{macro}-\cos\theta_Y)$,
with the macroscopic apparent contact angle $\theta^\mathrm{macro}$. 
Replacing in Eq.~\eqref{eq:CAHrow-a} the total macroscopic driving force
by the unbalanced Young's force we obtain 
\begin{equation}
	\gamma_{lv}(\cos\theta^\mathrm{macro}-\cos\theta_Y) 
	= - \frac{1}{L_y} \frac{\partial \Omega_0}{\partial x_{tl}}
	\bigg\rvert_{\emph{x}_{tl}=\emph{x}_{tl,st}}
\text{ .}
\label{eq:CAHrow}
\end{equation}
Equation~\eqref{eq:CAHrow} links the macroscopic contact angle  
at a given driving force $\Delta \mu$ 
to the microscopic defect force acting effectively on the triple line in
correspondence to its stable configurations identified by $x_{tl,st}(\Delta \mu)$.
The \emph{Supporting Information} provides an alternative route to
Eq.~\eqref{eq:CAHrow} based on Laplace's law.

A quantity of particular interest is the so-called
advancing contact angle $\theta_a$, which is  
the maximum apparent contact angle compatible 
with a static configuration during the advancing process.  
It is obtained by evaluating Eq.~\eqref{eq:CAHrow} 
at the maximum force the defects can exert upon opposing the advancing
process, yielding $\theta_{advancing}^\mathrm{macro}\equiv \theta_a$.
For an advancing barrier the maximum force is reached at $x_{tl,st}(\Delta \mu_{sp})$.
Similarly, the receding contact angle $\theta_r$ corresponds to the
maximum force (with opposite direction) the defects can exert upon
opposing the receding process. In order to obtain $\theta_r$ for a
receding barrier we likewise evaluate Eq.~\eqref{eq:CAHrow} at $x_{tl,st}(\Delta \mu_{sp})$,
but now $\Delta \mu_{sp}$ has a negative sign. For a pure advance barrier
the maximum force opposing the receding process is zero. Therefore, 
the RHS of Eq.~\eqref{eq:CAHrow} is zero so that $\cos\theta_r \equiv \cos\theta_Y$ for
this type of barrier. The above definitions of
$\theta_a$ and $\theta_r$ coincide with the usual experimental notion of
static contact angle hysteresis, which is measured just before 
the macroscopically defined contact line starts moving.      
For the posts discussed in the previous section 
(advance barrier, see Fig.~\ref{fig:hysteresis}A) the definitions of
$\theta_a$ and $\theta_r$, together with Eq.~\eqref{eq:CAHrow}, yield a very strong contact angle
hysteresis: $\cos\theta_r-\cos\theta_a=\cos\theta_Y
-\cos\theta_a=0.68$, corresponding to $\theta_a=121^\circ$ and
$\theta_r=81^\circ$. 
These two angles refer to a defect line density of $L_y^{-1}=0.057\;\sigma^{-1}$
corresponding to $L_y = 17.5\;\sigma$ and to the liquid-vapor interface advancing
simultaneously across all lined-up defects.

\subsection{Contact angle hysteresis induced by a random distribution of
defects}

We now consider a random, dilute distribution of identical
nanodefects. As discussed in the previous subsection, even in this case, away from the
walls the liquid-vapor interface attains its macroscopic shape
characterized by the apparent contact angle $\theta^\mathrm{macro}$ and
the  position $x_{tl}$ of the macroscopic triple line.
Therefore it is possible to establish a force balance between the unbalanced
Young's force on the macroscopic triple line and the forces provided by
the distribution of defects, analogous to Eq.~\eqref{eq:CAHrow}.
In order to provide an explicit equation, we first assume that the combined effect 
of the defects on a liquid front follows from linear superposition of the 
individual effects~\cite{joanny1984}. 
This additivity hypothesis, which is expected to be valid for dilute defect 
distributions, neglects the fact that also the presence of a neighboring defect
can alter the local shape of the triple line, thus affecting the metastable
configurations. Therefore, within this assumption, the free energy profiles and
the metastable configurations are the same for the quite distinct two cases of randomly distributed and aligned
defects. Actually, this is the content of a second hypothesis according to which the force
on the macroscopic straight triple line exerted by a single defect at a distance
$\Delta x_d\equiv x_d-x_{tl}$ 
from the triple line is the same as the one (for the same $\Delta x_d$) computed per
defect for a periodic row of defects sufficiently apart from each other. 
But in contrast to the above case of aligned defects, the force each defect exerts 
on the triple line depends on its distance $\Delta x_d$ from the macroscopic triple line, which is random. 
We therefore have to average the RHS of Eq.~\eqref{eq:CAHrow}
over the contribution of all defects adding to the force on a (straight) 
segment of the macroscopic triple line of length $L_y$. 

However, before performing this average in mathematical terms some
further remarks are needful
and we have to introduce a third hypothesis, which actually is related to the other two.
We first remark that in the absence of an external driving force the (mesoscopic) triple line
avoids crossing those regions where the forces are not negligible; these regions do not extend
much beyond the defects (i.e., less than an order of
magnitude of the defect size). Without this wiggling the
two hypotheses above would lead
to a non-zero mean defect force which is not balanced by an external driving force. 
These wiggles typically have a long wavelength, given by the average distance between the defects along
the triple line, and an amplitude which is small
and of the order of the linear extension of the defects, i.e.,
of nanometric size for the nanodefects discussed here. In order to
calculate the mean force
on the macroscopic triple line, we do not explicitly consider the
aforementioned wiggles of the (mesoscopic) triple line.
Instead, we average over the forces computed for various distances from the straight triple line, but
we exclude from the average those relative distances $\Delta x_d$ which would lead to locally
unstable configurations. This is the content of our third
hypothesis which by now is formulated,
taking into account an external driving force acting on the liquid-vapor interface. 

We are interested in the maximum absolute value of the total defect force on a triple line
segment, which can be provided by a random distribution of defects in an advance or a retreat experiment.   
We first focus on the advancing process across post-like defects
(exhibiting a barrier in advancing direction).
For a row of defects our calculations have shown that the defect force steadily increases
if the liquid-vapor interface is pushed against it until it reaches a maximum
at a driving force $\Delta \mu = \Delta \mu_{sp}$. 
In the case shown in Fig.~\ref{fig:hysteresis}B,
for $x_{tl}>x_{tl,st}(\Delta \mu_{sp})$ the position of the triple line becomes unstable and snaps 
over those parts of each defect in the straight row which are not
covered by liquid. We now assume that
\emph{locally} a similar behavior occurs for randomly distributed defects.
Once $\Delta x_{d}$ is larger than the critical
distance corresponding to the maximum force
which can be mustered by an individual defect, the triple line becomes locally unstable and slides over the 
still dry part of that specific defect. Accordingly, the defect forces corresponding
to these unstable positions do not contribute to the overall force average.
The average is performed only over forces corresponding to (meta)stable positions, 
which are characterized by $\Delta x_d =x_d-x_{tl,st}$. 
A relation to a driving force $\Delta \mu$ is not given, because only the total driving
force on the liquid front can be controlled, but not the force of a specific defect
acting on a segment of the macroscopic triple line. For an advancing process over defects
exhibiting a barrier in advance direction, $x_{tl,st}$ ranges from a
value $x_{tl}$ on the 
lower plateau in Fig.~\ref{fig:profiles} to $x_{tl,st} = x_{tl,sp}$,
characterizing the distance corresponding to the
maximum defect force. $x_{tl,sp}$ is identified
with  $x_{tl,st}(\Delta \mu_{sp})$ as determined from the computations for a row of defects. 
For a receding process over post-like defects, the defect forces are zero for all (meta)stable triple line positions.
For analyzing the case of defects providing a receding barrier one has to interchange
the role of advancing and receding processes in the discussion above. 

In order to provide quantitative results, we first introduce 
a length $l_d$ characterizing the range of distances from the triple
line within which the defect forces are non-negligible. 
The total number $M$ of defects interacting with the
segment $L_y$ of the triple line is $M = nL_yl_d$ with $n$ the areal number
density of the defects.  
The averaging is carried out by integrating over
all possible defect distances $\Delta x_d \equiv x_d - x_{tl}$
with equal weight $\mathrm d \Delta x_d/l_d$ (so that 
$\int_{-\mathit{l}_{d}/2}^{l_{d}/2} \; l_{d}^{-1} \; \mathrm d\Delta
x_d = 1$) or, equivalently, by integrating over the relative triple line position
$\Delta x_{tl} \equiv x_{tl} - x_d = -\Delta x_d$:      
\begin{align}
	\gamma_{lv}(\cos\theta^\mathrm{macro}-\cos\theta_Y) 
	&= - \frac{M}{l_{d}L_y}\int_{-l_{d}/2}^{l_{d}/2} \;
	\frac{\partial \Omega_0}{\partial x_{tl}}
	\bigg\rvert_{\emph{x}_{tl,st}} \;    
                       \mathrm{d} \Delta x_{tl} \nonumber\\
&= 
											 - n \left( \Omega_0(x_{tl,max}) -
											 \Omega_0(x_{tl,min}) \right)
\label{eq:CAHrandom}
\text{ .}
\end{align}
Without randomness Eq.~\eqref{eq:CAHrandom} reduces to
Eq.~\eqref{eq:CAHrow} with $M=1$.
In Eq.~\eqref{eq:CAHrandom} the notation should be interpreted such that integrands
are equal to zero for positions $\Delta x_{tl}$ which do not
belong to metastable positions $x_{tl,st}$ of the triple line along the advancing or the receding
processes. 
This also applies to the last equation, in which we evaluate
$\Omega_0$ at the maximum and the minimum values of $x_{tl,sp}$ 
($x_{tl,max}$ and $x_{tl,min}$, respectively).
For instance, the advancing process over post-like defects is
characterized by $x_{tl,max}^\mathrm{a} = x_{tl,sp}$, while
$x_{tl,min}^\mathrm{a}$ corresponds to some position $x_{tl}$ along the lower plateau in
the free energy profile, e.g., $x_{tl,v}$ (Fig.~\ref{fig:hysteresis}A).
Evaluating Eq.~\eqref{eq:CAHrandom} over the whole metastable interval 
$x_{tl,min}^\mathrm{a}<x_{tl,st}<x_{tl,max}^\mathrm{a}$ of the advancing process yields the
advancing contact angle $\theta^\mathrm{macro}_{advancing}$.
For the receding process across the same kind of defects there are no
metastable configurations of the system ($x_{tl,st}\in \emptyset$) and therefore the
integral in Eq.~\eqref{eq:CAHrandom} vanishes, yielding $\cos\theta^\mathrm{macro}_{receding}-\cos\theta_Y=0$.

As shown in Fig.~\ref{fig:hysteresis}A, for post-like defects
$x_{tl,max}^\mathrm{a} (\Delta \mu_{sp})\equiv x_{tl,sp}$
and $x_{tl,min}^\mathrm{a} (0)\equiv x_{tl,v}$, yielding
$\gamma_{lv}(\cos\theta_a-\cos\theta_Y)=-n \Delta\Omega_0(x_{tl,sp})$
where the maximum contact angle compatible with Eq.~\eqref{eq:CAHrandom} coincides with
the notion of the advancing contact angle introduced before, $\theta^\mathrm{macro}_{advancing}\equiv
\theta_a$, and $\Delta \Omega_0 (x_{tl,sp})=
\Omega_0(x_{tl,sp})-\Omega_0(x_{tl,v})$;
for the receding process, instead,
$\theta^\mathrm{macro}_{receding}\equiv \theta_r$ and $\cos\theta_r-\cos\theta_Y=0$.
Combining the previous two expressions, we obtain an estimate for the contact angle
hysteresis:
\begin{equation}
	\cos\theta_r-\cos\theta_a=n\frac{\Delta\Omega_0(x_{tl,sp})}{\gamma_{lv}}
	\label{eq:CAH}
\text{ .}
\end{equation}
Equation~\eqref{eq:CAH} is valid also for retreat barriers
(Fig.~\ref{fig:hysteresis}B) for which, however,
$\cos\theta_a-\cos\theta_Y=0$ and
$\gamma_{lv}(\cos\theta_r-\cos\theta_Y)=n \Delta\Omega_0(x_{tl,sp})$.

The values of $\Delta\Omega_0(x_{tl,sp})$ in Table~\ref{tab:barriers}
together with Eq.~\eqref{eq:CAH} allow us to 
determine which
density of nanodefects of the post kind, with the characteristics
chosen in the present investigations,
is needed in order to induce macroscopically
relevant pinning. 
As a typical measurable difference in cosines we adopt
$\cos \theta_r -\cos\theta_a=0.25$, use the surface tension of our simple
liquid $\gamma_{lv}=0.406\;\varepsilon \sigma^{-2}$, and assume
$\sigma=0.3$~nm, which yields
for the post defect a projected area $A_d=0.56$~nm$^2$.  Since 
$\Delta\Omega_0(x_{tl,sp})\approx 10$~$\varepsilon$ for
a single post defect,
from Eq.~\eqref{eq:CAH} we obtain
$n=0.11$~nm$^{-2}$, corresponding to a surface coverage $n \, A_d$ of ca.
$6\%$. 
This simple calculation for a post of size
$0.75\;\mathrm{nm}\times0.75\;\mathrm{nm}\times0.75\;\mathrm{nm}$ and
Young contact angle $\theta_Y\approx125^\circ$
suggests that even dilute chemical and topographical defects of sub-nanometer scale can be
responsible for the pinning phenomena associated with nanobubble
stability~\cite{weijs2013}. For comparison, the contact angle hysteresis
of the substrate employed in recent experiments on nanobubbles
\cite{zhang2013} was $\cos\theta_r-\cos\theta_a=0.25$ for a slightly
larger roughness.

\section*{Conclusions}
We have studied a liquid wedge advancing or receding across a single
chemical or topographical heterogeneity of nanometric size.
We have devised a method which combines string calculations with microscopic
classical density functional theory in order to determine free energy profiles for 
overcoming such obstacles. This approach has allowed us to take into account 
specific microscopic effects, which have turned out to be significant
on the nanoscale. At liquid-vapor coexistence the free energy profiles 
exhibit as function of the position of the triple line a sigmoidal 
shape with two plateaus which correspond to the heterogeneity being exposed solely to vapor
or completely covered with liquid, respectively.
For the topographical and chemical protrusion studied, the macroscopic estimate for the 
difference of these plateau values is half of the actual microscopic
result.  Based on the computed free energy 
profiles we have discussed thermally activated depinning of the three
phase contact line in the absence of driving forces.       
We have shown that wetting hysteresis
originates from the occurrence of a metastable ``pinned'' state within a
certain range of chemical potentials. Forcing the system towards and off
two-phase coexistence leads to qualitatively distinct advance and
retreat branches.  
The defects considered here are weak in terms of the classification of
Joanny and de Gennes; however, some of them nonetheless give rise to wetting
hysteresis.
Our estimates of the macroscopic effects of a distribution of nanoscale
defects indicate that a $6\%$ surface coverage of lyophobic topographical defects
with a lateral extension smaller of $1$~nm is sufficient to induce a measurably large
contact angle hysteresis, i.e., $\cos\theta_r-\cos\theta_a=0.25$. These figures match
well with those obtained from recent nanobubbles experiments. 
Our results suggest that this provides a mechanism for contact line pinning
which is needed to explain the unexpected long lifetime of the nanobubbles.
Finally, the techniques introduced here and the proposed framework for discussing
wetting hysteresis have laid a common probabilistic ground for understanding deterministic
and thermally activated pinning of three-phase contact lines. Accordingly, this analysis is expected to clear the way for
future detailed microscopic studies of activated processes in fluids at small scales.

%%%%%%%%%%%%%%%%%%%%%%%%%%%%%%%%%%%%%%%%%%%%%%%%%%%%%%%%%%%%%%%%%%%%%

%%%%%%%%%%%%%%%%%%%%%%%%%%%%%%%%%%%%%%%%%%%%%%%%%%%%%%%%%%%%%%%%%%%%%

\paragraph{ACKNOWLEDGMENTS.}
A.G. thanks S. Meloni for thoughtful discussions. 
The authors are grateful to A. Checco and T. Ondar\c{c}uhu for
useful comments.

%%%%%%%%%%%%%%%%%%%%%%%%%%%%%%%%%%%%%%%%%%%%%%%%%%%%%%%%%%%%%%%%%%%%%

\bibliographystyle{pnas2009}

%%%%%%%%%%%%%%%%%%%%%%%%%%%%%%%%%%%%%%%%%%%%%%%%%%%%%%%%%%%%%%%%%%%%%

\newpage
\clearpage
\appendix
\renewcommand{\thefigure}{S\arabic{figure}}
\renewcommand{\thesection}{S\arabic{section}}
\setcounter{section}{0}    

\section*{Supporting Information}

\section{Density functional theory}

\subsection{Fluid-fluid and fluid-wall interactions}
In the present study we use the fundamental measure theory (FMT)
due to Rosenfeld in order to determine the molecular details of the
fluid. This classical density functional theory provides a non-local functional $F_{FMT}[\rho]$
which accounts for the repulsive part of the interaction
between the fluid particles, taken to be a \emph{h}ard-\emph{s}phere pair potential:
$\Phi_{HS}=\infty$ 
if the distance between the centers of two fluid particles is smaller than the
particle diameter $\sigma$, zero otherwise. The \emph{att}ractive
part $\Phi_{att}$ of the interaction between the fluid particles is included
perturbatively by applying the spirit of the Weeks-Chandler-Andersen
theory \cite{weeks1971} to the Lennard-Jones potential:  
\begin{eqnarray}
        \Phi _{\mathrm{att}}(r) & = &
                - \, \varepsilon \, \Theta\left( 2^{1/6}\sigma - r \right) +  \nonumber \\
       &   &	4 \, \varepsilon \left[\left(\frac{\sigma}{r}\right)^{12} -
		\left(\frac{\sigma}{r}\right)^{6}\right] 
                  \Theta\left( r - 2^{1/6}\sigma \right)       
	\text{ ,}
	\label{eq:potential}
\end{eqnarray}
where $r$ is the distance between the centers of a pair of particles, 
$\varepsilon$ is the
interaction strength, $\sigma$ corresponds to the particle diameter, and
$\Theta$ is the Heaviside function. In the following, $\varepsilon$ sets
the energy scale. 
In the calculations a cutoff of $2.5\;\sigma$ is employed for $\Phi_{att}(\bm r)$.
Concerning the detailed form of $F_{FMT}[\rho]$ and the calculation details see
Refs.~\cite{roth2010,singh2015}.

%Fig1
\begin{figure}[b!]
	\centering
	\includegraphics[width=0.8\textwidth]{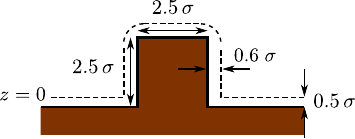}
	\caption{
		Schematic drawing of the wall with a post defect, here a cube. 
		The wall is defined as the semi-infinite region in brown occupied by the centers of
		the wall particles.
		The dashed lines correspond to the distance of closest approach to the wall for the
		centers of the fluid particles; 
		beneath these
		lines $V(\bm r)=100\,\varepsilon$ and
		the number density of the fluid is de facto zero.  The plane
		$z=0$ is located at the lower horizontal dashed line as indicated.
	\label{fig:walldef}}
\end{figure}

The external substrate potential $V(\bm r)$ in Eq.~\eqref{eq:omega} of the main text is  
expressed as the sum of a repulsive contribution $V_{\mathrm{rep}}(\bm
r)$ and an attractive one $V_{\mathrm{att}}(\bm r)$:
$V(\bm r) = V_{\mathrm{rep}}(\bm r) + V_{\mathrm{att}}(\bm r)$. 
We account for $V_{\mathrm{rep}}(\bm r)$ in terms
of a hard-sphere repulsion, chosen such that the distance of
closest approach of the 
centers of a fluid particle and a wall particle with diameter
$\sigma_w$ is $\sigma_{wall} = (\sigma + \sigma_w)/2 $.
$V_{\mathrm{rep}}(\bm r) = \infty$ if the distance between the center 
of a fluid particle and a wall particle is smaller than $\sigma_{wall}$
(the region beneath the dashed lines in
Fig.~\ref{fig:walldef}), $V_{\mathrm{rep}}(\bm r) = 0$ otherwise.
For the bottom part of the wall we have used the limit
$\sigma_w/\sigma \rightarrow 0$, i.e., $\sigma_{wall} = \sigma /2$.
Around the post, instead, $\sigma_{wall} = 0.6\;\sigma$ has been chosen.
In the presence of a non-zero and fixed $V_{\mathrm{att}}(\bm r)$
a larger value of $\sigma_{wall}$ around the post corresponds to a more
lyophobic behavior; this is due to the fact that repulsive interactions
overrule attractive ones in a more extended region close to the defect,
see Fig.~\ref{fig:walldef}.
$V_{\mathrm{att}}(\bm r)$  is taken as the linear
superposition of the attractive part $\Phi_w$ of the Lennard-Jones potential
between the fluid particles and the particles forming the wall with a
number density $\rho_w$:
\begin{equation}
	V_{\mathrm{att}}(\bm r) = \int_{D_w} \mathrm{d}^3 r\, ' \; \rho_w(\bm r\, ') \; \Phi_w(\bm r -\bm r\, ') \text{ .}
	\label{eq:wallpot}
\end{equation}
Here the fluid-wall pair potential $\Phi_w(\bm r -\bm r\,')$, which
depends on the distance between the positions $\bm r$ and $\bm r\,'$ of the fluid and wall
particles, respectively, is integrated over the domain $D_w$ occupied by
the wall; as stated above $\Phi_w(\bm r -\bm r\, ')=-4\,\varepsilon_w(\bm r\, ')
(\sigma_w/|\bm r -\bm r\, '|)^6$.  
The strength of the wall potential is determined by the parameter combination 
$u_{w}(\bm r\, ') \equiv (2\pi/3)\, n_{w}(\bm r\, ')\,\varepsilon_{w}(\bm r\, ')$ 
where $n_{w} = \rho_w \sigma_{w}^3$ is the number of wall particles in
the volume $\sigma_{w}^3$.
The specific parameter $u_{w}$, subsequently called
wall energy, has been introduced because it characterizes the attractive potential
$V_{\mathrm{att}}(d) = -u_{w}/d^3$ acting on a fluid particle
with its center at a distance $d$
from the planar surface of a half-space of 
continuously and homogeneously distributed wall particles
of a certain kind.  
More details can be found in Ref.~\cite{singh2015}.
In order to model a wetting contrast between the wall and a chemical
blemish, the
wall material underneath the corresponding surface area is replaced, up to a certain depth,
by material characterized by a wall energy $u_{w}(\bm r)$ which is distinct from the
wall energy $u_w$ characterizing the bulk of the wall. 
Unlike in macroscopic
models, such as the ``mesa'' defects considered by Joanny and de Gennes 
\cite{joanny1984} for which the wetting properties of the substrate
vary step-like, here the wall potential $V(\bm r)$ varies smoothly even
where $u_w$ varies discontinuously in the lateral spatial directions.

The chemical defects are obtained by starting from a homogeneous wall
occupying the lower half-space with $u_w=3\,\varepsilon$
(see above). Then a parallelepiped of
size $\Delta x\times \Delta y\times \Delta z= 2.5\,\sigma \times
2.5\,\sigma \times 3\,\sigma$ is carved
out at the surface of the semi-infinite wall and
replaced by one with a different value of $u_w$. The resulting heterogeneous wall is still
planar ($\sigma_{wall}$ is constant and equal to $\sigma/2$ over the entire
wall) but with an inclusion of a different material
and of depth $3\,\sigma$. 
We classify these chemical defects based on the contact angles measured
for homogeneous semi-infinite walls having the same $u_w$ as the
inclusion:
$u_w=2\,\varepsilon$ corresponds to a lyophobic
substrate with $\theta_Y=124.85^\circ$, 
$u_w=3.5\,\varepsilon$ corresponds to a substrate which
is partially wet with $\theta_Y=51.13^\circ$, 
and $u_w=4.0\,\varepsilon$ corresponds to a substrate which
is completely wet.

The system used in the present analysis is shown in Fig.~1B of the
main text. The density is initialized with the bulk liquid and vapor
densities at two-phase coexistence at the right and the left side of the
computational box, respectively. At the right and left ends of the
computational box we have chosen reflecting boundary conditions (the
$y-z$ planes at both ends are mirror symmetry planes); this is virtually
equivalent to having capillary liquid boundary conditions on the right
and capillary vapor boundary conditions on the left, as long as the
liquid-vapor interface and the surface defects introduced in the next
section are sufficiently far away from the box boundaries.  The
remaining two boundaries at the front and at the rear exhibit periodic
boundary conditions.  
The domain is discretized with $20$ points per particle diameter
$\sigma$, such that the actual computational domain consists of
$650\times350\times300=68\,250\,000$ points.

\subsection{Wall potential for the post defect}

%Fig2
\begin{figure}[h]
	\centering
	\includegraphics[width=0.9\textwidth]{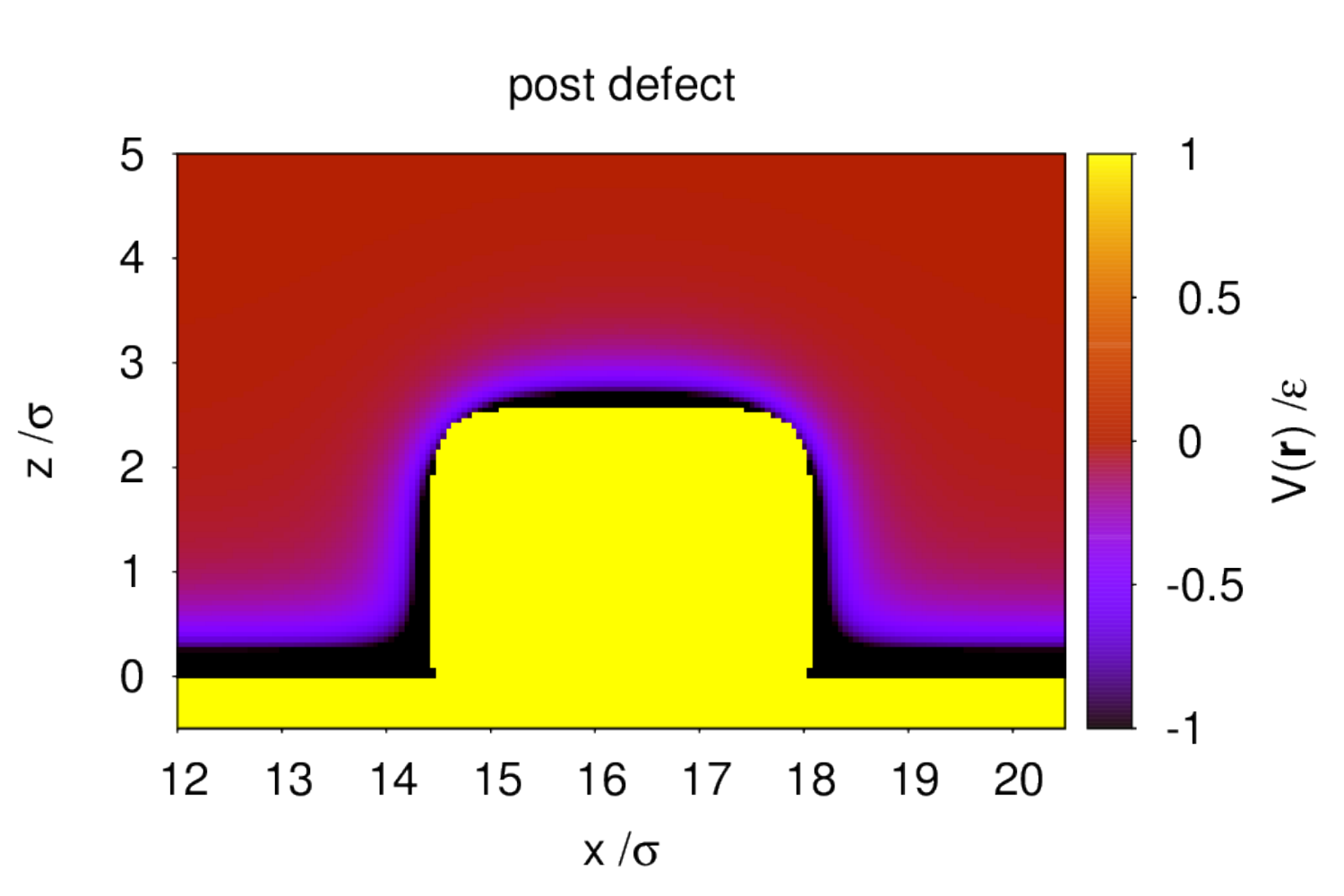}
	\caption{Wall potential $V(\bm r)$ in units of $\varepsilon$ in the vicinity of the post
	defect within a vertical cut $y=8.75\sigma$ through its center at
	$(x,y,z)=(x=16.25\sigma,y=8.75\sigma,z=1.25\sigma)$. The post is a
	cube of size 
	$\Delta x \times \Delta y \times \Delta z =
	2.5\sigma\times2.5\sigma\times2.5\sigma$ and the  computational
	box has the size $\Delta x \times \Delta y \times \Delta z = 32.5\sigma \times
	17.5\sigma \times 16\sigma$. The yellow domain, where $V(\bm
	r)$ is positive, represents the region inaccessible to the fluid
	due to harshly repulsive interactions.
	\label{fig:wall}}
\end{figure}
In Fig.~\ref{fig:wall} we show the wall potential $V(\bm r)$ for the
post defect introduced in the main text. Within the yellow domain one has
repulsive interactions between the fluid and the wall consisting of a
hard core with $V(\bm r)=\infty$. In practice, the potential has been
set to the large but finite value $V(\bm r)=100\,\varepsilon$. The hard core of the lower wall is
located at $z/\sigma=0$. The top corners of the post are
not perfectly smooth because for numerical reasons the potential is defined on a discrete
lattice and due to the way we test the overlap between the hard wall
and the hard core of the fluid particles.
Figure~\ref{fig:wall} demonstrates that, although the bottom wall and
the post share the same attractive interaction strength
$u_\mathrm{w}$, the potential next to the spatial regions dominated by
strongly repulsive interactions (defining depletion zones within which
the liquid densities are zero) is very different at the bottom wall
(where it reaches $-4\,\varepsilon$) or near the post (where it reaches $-2\,\varepsilon$).
This difference is partially due to the attractive part of the interaction being 
reduced by the enlarged depletion zone ($\sigma_{wall}=0.6\;\sigma$) introduced in our model of
the post as compared to the planar bottom part of the wall ($\sigma_{wall}=0.5\;\sigma$). A second reason is the fact
that the wall particles at the top of the post
have fewer close neighbors than the particles at the bottom surface of the
wall. The combination of these two effects leads to an effectively
lyophobic behavior of the post (see $\delta \Omega_{post}>0$ in
Table~1 and Fig.~2B in the main text).

%%%%%%%%%%%%%%%%%%%%%%%%%%%%%%%%%%%%%%%%%%%%%%%%%%%%%%%%%%%%%%%%%%%%%

\section{Combining microscopic DFT with the string method}
\label{sec:algorithm}
 
Our calculations combine classical density functional
theory (DFT) with the string method as follows.

%Code
The DFT code we are using implements the fundamental measure theory (FMT)
functional for the repulsive part of the pair interaction between the
fluid particles. The same
code also computes, via a separate algorithm, the wall potentials (as
the one shown, \emph{e.g.}, in Fig.~\ref{fig:wall}). The free minimization (without
string) of the grand potential defined in Eq.~\eqref{eq:omega} of the main text is
performed via the Picard iteration technique. 

%Definition of the string
The string $\rho(\bm r, \tau)$ is defined as a sequence of number
density configurations spanning the entire activated process  -- in the present case the
advance or retreat of a liquid wedge across a defect -- from an
initial state $A$ to a final one $B$. The variable $\tau$ is the arc
``length'' computed along the string:
\begin{equation}
\tau = \int_{s_A}^{s_L} \mathrm d s \; \sqrt{\int_{D_c}\; \mathrm d^3 r \;
\left(\frac{\partial \rho(\bm r, s)}{\partial s} \right)^2}  
\text{ ,}
\label{eq:arclength}
\end{equation}
where $A$ is the initial state, $D_c$ is the computational domain, $L$ is a generic point of the string
with $s_A<s_L<s_B$, and $s$ is an arbitrary parametrization of the string.  

%Initial conditions for the string
We construct the initial string by a rough free minimization ($\approx 300$
iteration steps) of $20$ to $22$ suitably selected  independent fluid configurations -- the so-called
images. These initial images are computed starting from distinct 
initial conditions which consist of a vertical liquid-vapor interface
separating the liquid domain to the right and the vapor one to the left.
Each configuration corresponds to a different filling level of the
channel, spanning it from the far left of the defect to its far right. After
the pre-equilibration, the initial string consists of liquid wedges like
the one in Fig.~1B of the main text translated along the horizontal
$x$-axis.

The most probable path defined in Eq.~\eqref{eq:MFEP} of the main text
can be computed via the following pseudodynamics for the images:
\begin{equation}
 \frac{\partial \rho(\bm r, \tau, t)}{\partial t} = 
- \left(\frac{\delta \Omega[\rho]}{\delta \rho (\bm
  r,\tau,t)}\right)_\perp
\text{ ,}
\label{eq:dynamics1}
\end{equation}
which at stationarity reduces to Eq.~\eqref{eq:MFEP} of the main
text. 

One way to compute the functional derivative perpendicular to the string
(denoted as $\perp$ in Eq.~\eqref{eq:dynamics1}) is to subtract from $\delta
\Omega[\rho]/\delta \rho (\bm r,\tau,t)$ the component tangential to the
string: 
\begin{equation}
 \frac{\partial \rho(\bm r, \tau, t)}{\partial t} = 
- \frac{\delta \Omega[\rho]}{\delta \rho (\bm r,\tau,t)} 
+ \lambda\, \hat T(\bm r,\tau,t) \text{ ,}
\label{eq:dynamics}
\end{equation}
where, using the $\mathcal{L}^2$-norm, the tangent $\hat T(\bm r,\tau,t)$ to the string is defined
 as 
$\hat T(\bm r,\tau,t)  \allowbreak \equiv (\partial \rho(\bm r,\tau, t)/\partial
\tau)/\|\partial \rho(\bm r,\tau, t)/\partial  \tau \|$ and  $\lambda \equiv \langle\delta
\Omega[\rho]/\delta \rho (\bm r,\tau,t),\hat T(\bm r,\tau,t)\rangle$,
with $\langle\; ,\,\rangle$ indicating the inner product defined as
$\langle f, g \rangle \equiv \int_{D_c} \mathrm d^3 r\; f(r) g(r)$
within the computational domain $D_c$  (see, e.g., Fig.~2 of the main text).
In the dynamics for the string described by Eq.~\eqref{eq:dynamics},
the tangential component of $\delta \Omega[\rho]/\delta \rho (\bm r,\tau,t)$ does not change the string
itself but it affects its parametrization $\tau$, displacing the
images along it (which is a general result for curve dynamics,
see Ref.~\cite{e2007}).
In other words, the term $\lambda \hat T (\bm r,\tau,t)$ prevents 
the images along the string from collapsing into the closest minima. 

%
%Fig3
\begin{figure}[h]
	\centering
	\includegraphics[width=0.85\textwidth]{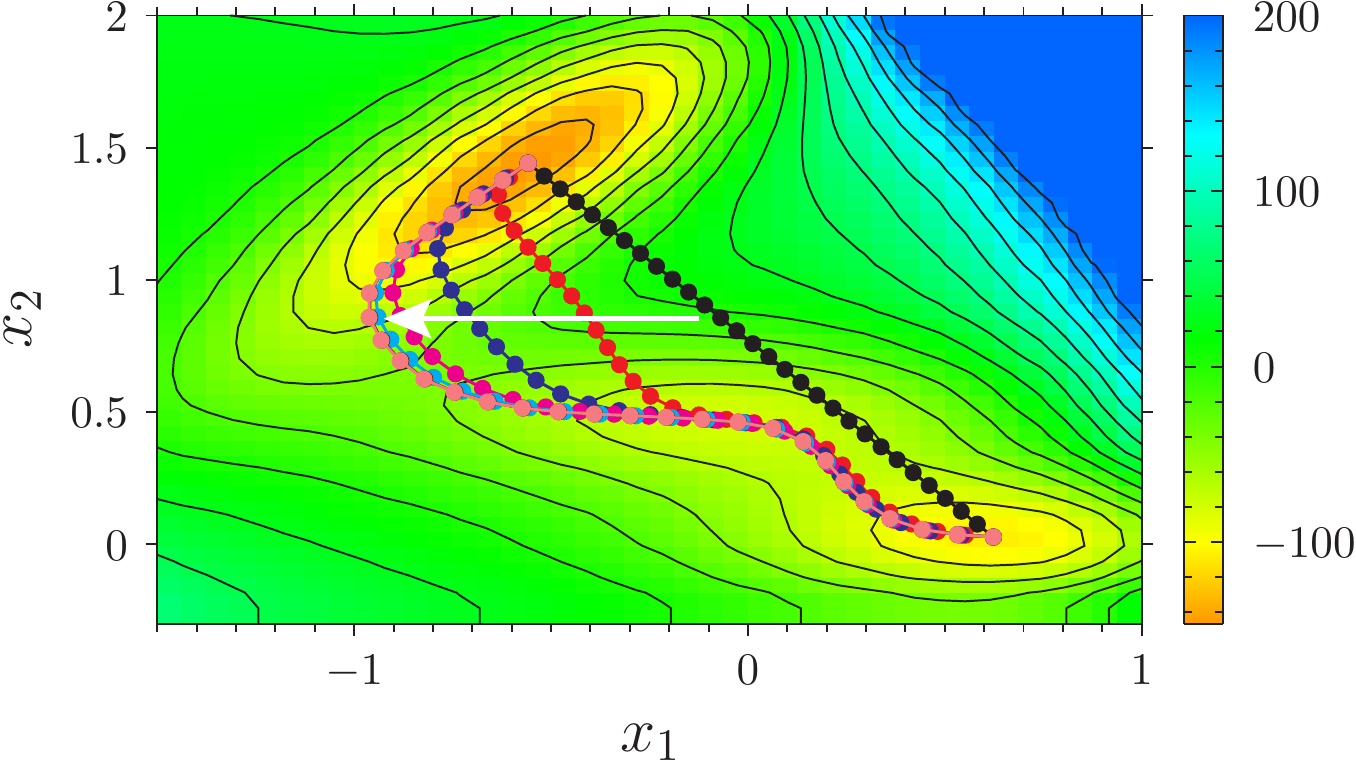}
	\caption{
			Illustration of the string method on a two-dimensional model energy landscape,
			generated by the so-called M\"uller potential used in Ref.~\cite{e2007}, which depends on
			two dimensionless parameters $x_1$ and $x_2$.
			Color code and contour lines provide the value of the potential in
			dimensionless units. The dots denote the
			images in which the string is discretized. The white arrow
			indicates the direction in which the images evolve according to
			Eq.~\eqref{eq:dynamics}, starting from
			their initial position (black) and ending at the converged one
			(pink). 
			The two ends of the string are initialized at the two minima of
			interest.
\label{fig:string}}
\end{figure}

In order to ensure that the images maintain a particular parametrization
during the evolution, in Eq.~\eqref{eq:dynamics} we
consider explicitly an additional term. 
Here we employ the equal arc length parametrization 
which guarantees that two contiguous images remain at the same normalized
distance along the string during the evolution. For the continuous
string this constraint reads $\|\partial \rho(\bm r,\tau, t)/\partial
\tau \|=\mathrm{const} $, which can be enforced by Lagrange
multipliers. In practice the procedure consists in adding the term
$\bar\lambda\, (\partial \rho(\bm r,\tau, t)/\partial
\tau)/\|\partial \rho(\bm r,\tau, t)/\partial  \tau \| \equiv \bar
\lambda\, \hat T(\bm r,\tau,t)$ to Eq.~\eqref{eq:dynamics}, see
Ref.~\cite{e2007}. This additional term does not change the
functional form of Eq.~\eqref{eq:dynamics} provided that $\lambda +\bar
\lambda$ is substituted for $\lambda$. This additional
Lagrange multiplier enforces a particular parametrization of the string
and thus Eq.~\eqref{eq:dynamics} can be solved via the two-step
procedure described in the following \cite{e2007}.
The string algorithm just described is illustrated in
Fig.~\ref{fig:string} for the so-called M\"uller potential
which depends on two dimensionless parameters $x_1$ and $x_2$ \cite{e2007}.
For the discretized string actually employed in the present calculations, the
normalized distance $d$ between two generic images $i$ and $i+1$ is
considered, i.e., $d= \sqrt{\int\; \mathrm d^3 r \;(\rho(\bm r, \tau_i)
-\rho(\bm r, \tau_{i+1}))^2}/\tau_{tot}$, where $\tau_{tot}$ is the
total length of the string computed from Eq.~\eqref{eq:arclength} taking
$s_L\equiv s_B$.

Following again Ref.~\cite{e2007} we use a time splitting procedure for the
two terms on the right-hand side of Eq.~\eqref{eq:dynamics}. First, the
unrestrained dynamics 
\begin{equation}
 \frac{\partial \rho(\bm r, \tau, t)}{\partial t} = 
- \frac{\delta \Omega[\rho]}{\delta \rho (\bm r,\tau,t)} 
\label{eq:dynamics3}
\end{equation}
is solved for each image. Second, the equal arc length parametrization
is enforced, not by computing explicitly $\lambda+\bar \lambda$ and
$\hat T(\bm r,\tau, t)$, but by a simpler interpolation step, in which
the images are redistributed along the interpolated string so that
$d=1/(\mathcal M-1)$ where $d$ is the normalized distance between any
two contiguous images and $\mathcal M$ is the total number of images.  
The significant advantage of this two-step algorithm is
that for the first step (Eq.~\eqref{eq:dynamics3}) the standard free energy
minimization procedure of DFT can be used, such as the Picard iteration
implemented in our code.
For the second step, 
with cubic splines we	interpolate the values of the number densities 
$\rho(\bm r, \tau_i,t)$ along the string using as independent
variable the arc length $\tau_i$ of the string up to the \emph{i}-th
image; $\tau_i$ is computed from	Eq.~\eqref{eq:arclength} taking
$s_L=s_i$ with the image index taking the values
$i=0,\ldots,\mathcal M-1$. The new, interpolated string is
constructed by computing the number densities $\rho^{new}(\bm r,
\tau^{new}_i,t)$ at the points $\tau_i^{new}$ which are uniformly
distributed along the string: $\tau_i^{new}\equiv \tau_{tot} \, i
/(\mathcal M -1)$, where the total length of the string is $\tau_{tot}\equiv
\tau_{\mathcal{M}-1}$.
In sum, the hybrid string - DFT algorithm consists of $\mathcal
M$ short, independent free minimizations of the grand potential (one for
each image) alternated with an interpolation step.

Evidently, the
string method seeks the most probable path in a space with the
same dimensionality as the discretized number density, \emph{i.e.}, ca.
$10^8$. In this scheme, no \emph{a priori} assumption is made on what are the
relevant parameters to describe the activated process (the so-called
collective variables in the language of studies of rare events).

%%%%%%%%%%%%%%%%%%%%%%%%%%%%%%%%%%%%%%%%%%%%%%%%%%%%%%%%%%%%%%%%%%%%%

\section{Parametrization of the string}

%Fig4
\begin{figure}[h]
	\centering
	\includegraphics[width=0.7\textwidth]{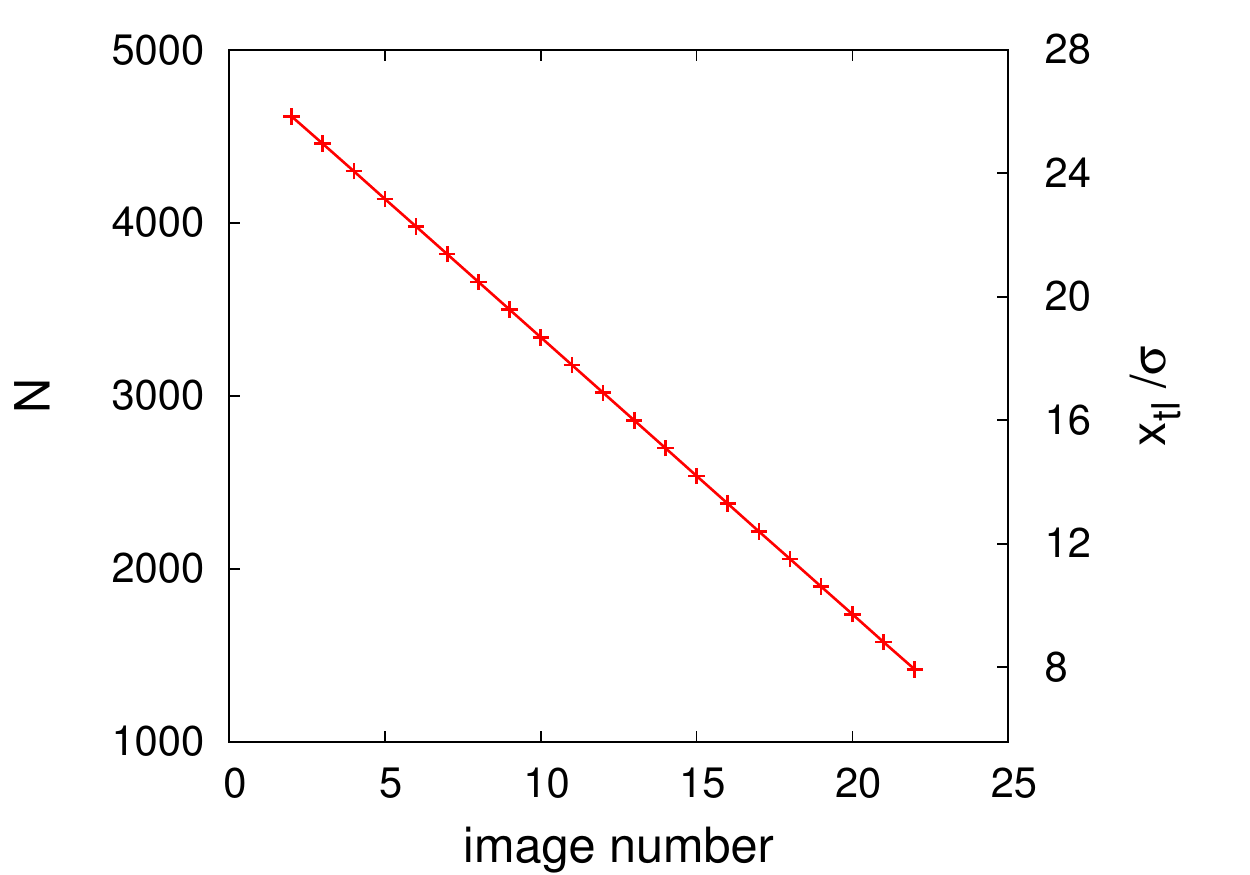}
	\caption{Number of particles $N$ in the computational box of
	size $\Delta x \times \Delta y \times \Delta z =32.5\sigma \times 17.5\sigma \times 16\sigma$ for the images of
	the string in the case of the defect-free channel. The position
	$x_{tl}= N /(L_y L_z (\rho_l -\rho_v))$ of the triple line  is 
	shown, too, on the right vertical axis. The cases
	with defects also exhibit linear relationships between the image
	number and $N$ or $x_{tl}$.  \label{fig:imageN}}
\end{figure}

Based on the discussion in Sec.~\ref{sec:algorithm} above, the natural
parametrization of the string is the arc length $\tau$ defined in
Eq.~\eqref{eq:arclength}. Since, however, the images are equally spaced along the
string, $\tau$ and the image number are both valid parametrizations of
the string.

As shown in Fig.~\ref{fig:imageN}, it is also possible to verify
\emph{a posteriori} that the most probable path can be parametrized by the total
number  $N\equiv\int\mathrm d^3 r \, \rho(\bm r)$ of particles in the
computational box. The dependence of $N$ on the image number is apparently
linear.
This parameter is easy to compute and has a direct physical meaning,
related to the progressive filling of the channel during the advance of
the liquid wedge. Furthermore, the parameter $N$ is also useful for
directly applying Eq.~\eqref{eq:shift} of the main text as has been done,
\emph{e.g.}, in  Fig.~\ref{fig:comparison}. 

The variable $N$ can also be related to the position $x_{tl}$ of the triple line
 via $x_{tl}= N /(L_y L_z (\rho_l -\rho_v))$ (see Fig.~1B and
the main text). Clearly, $x_{tl}$ maintains a linear dependence on  the
image number (Fig.~\ref{fig:imageN}). The position of the triple line
can be monitored experimentally (see, e.g., Ref.~\cite{delmas2011});
accordingly, $x_{tl}$ is used throughout the main text as well as for the free
energy profiles presented in the following section.

%%%%%%%%%%%%%%%%%%%%%%%%%%%%%%%%%%%%%%%%%%%%%%%%%%%%%%%%%%%%%%%%%%%%%

\section{Lyophobic and partially wet patches}

%Fig5
\begin{figure}[hp]
	\centering
	\includegraphics[width=0.65\textwidth]{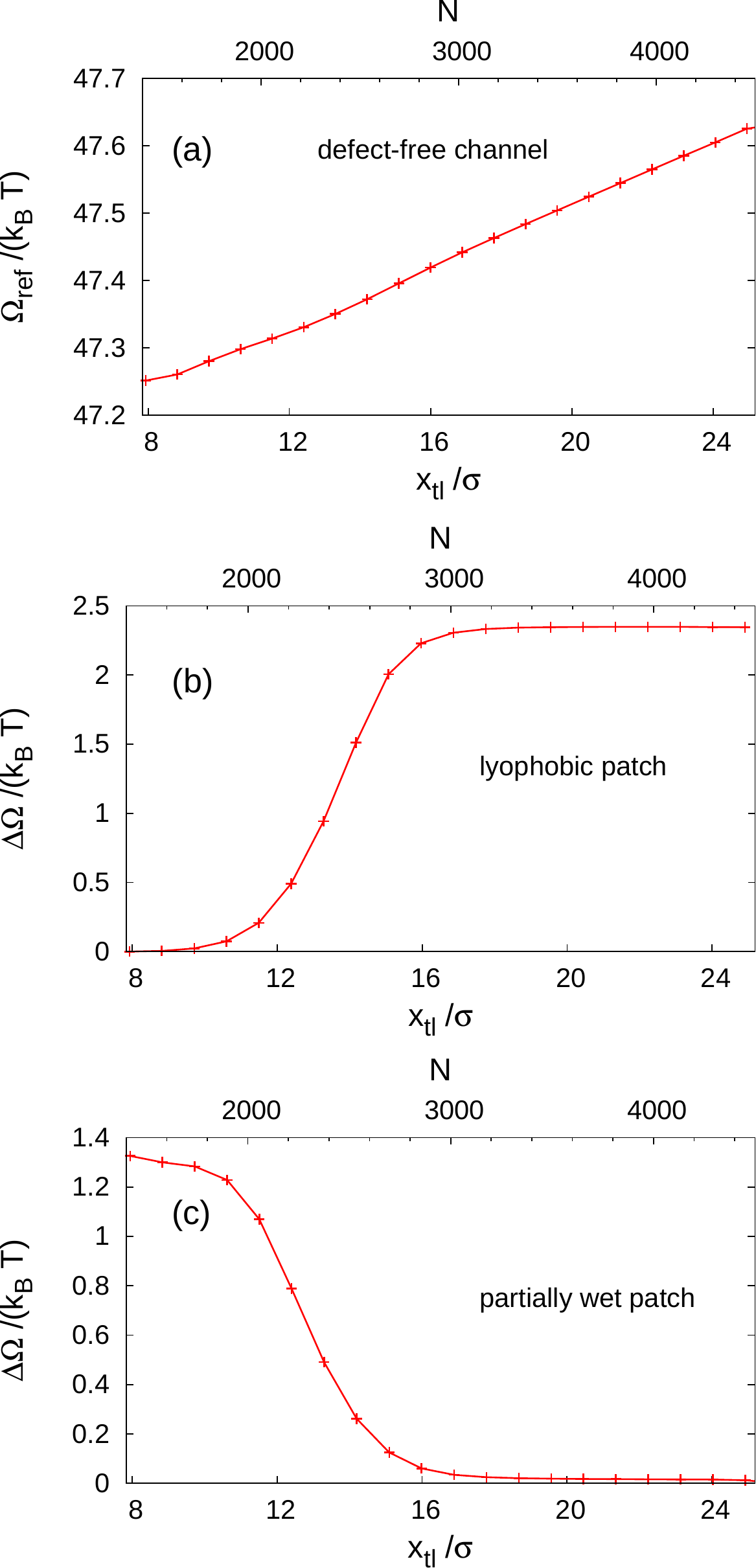}
	\caption{(a) Reference free energy, taken to be that of a liquid
	wedge advancing or receding in a defect-free channel at
  bulk two-phase coexistence $\mu=\mu_0(T)$. The excess free energy
	$\Delta \Omega$ at two-phase coexistence is shown
	for a lyophobic patch (b, $u_w/\varepsilon=2$) and for a partially
	wet patch (c, $u_w/\varepsilon=3.5$).
	\label{fig:profiles}}
\end{figure}

In this section we report the free energy profiles for the defect-free
channel and for the kind of defects not discussed explicitly in the main
text, \emph{i.e.}, the lyophobic and partially wet patches. Concerning
the related transition paths, see, c.f., the Videos S$1$-$2$ supplied in 
Sec.~\ref{sec:videos}.

Figure~\ref{fig:profiles}(a) shows the free energy profile related to the
filling of the defect-free channel. The profile is not
constant but it exhibits to a large extent a linear dependence on
$x_{tl}$.
This tells that, in spite of the
judicious choice of the substrate strengths $u_w$, for the channel under
study bulk coexistence $\Delta \mu=0$ does not exactly correspond to
coexistence between capillary liquid and capillary vapor. This
discrepancy is due to the imperfect matching of the lower
and upper contact angles $\theta_Y^{low}+\theta_Y^{up}=180.18^\circ$ and
possibly also due to microscopic effects such as liquid layering.
The thermodynamic force resulting from the small slope
of the free energy profile for the defect-free channel is in practice
very small. After relaxing to the configuration shown in Fig.~1B of
the main text, the system indeed moves towards the global free energy
minimum but extremely slowly: even after iterating several hundreds of
times the unrestrained minimization algorithm, the change in the number
density is negligible.

The images for small and large $x_{tl}$ do not lie on a perfect straight line; it can be verified from the
corresponding number density configurations that this deviation is due to the
interaction of the liquid-vapor interface with the reflecting boundaries.
Accordingly, these images have not been used for computing the free energy
barriers, the locations of the minima, etc. 

The defect-free free energy profile $\Omega_{\mathrm{ref}}(x_{tl})$ in
Fig.~\ref{fig:profiles}(a) is used as
a reference for computing the profiles in cases in which a defect is
present; from the free energy profiles $\Omega(x_{tl})$ computed by the string method we 
subtract $\Omega_{\mathrm{ref}}(x_{tl})$ at the corresponding values of
$x_{tl}$
yielding the excess free energy profiles $\Delta \Omega(x_{tl})\equiv
\Omega(x_{tl})-\Omega_\mathrm{ref}(x_{tl})$ reported in
Figs.~\ref{fig:profiles}(b) and (c) as well as in Fig.~3 of the main text. With this
definition, the excess free energy profiles indeed correspond to those
of an unbounded liquid wedge advancing or receding across a defect.

Figures~\ref{fig:profiles}(b) and (c) show the excess free energy profiles
for a chemical defect of lyophobic character ($u_w/\varepsilon=2$) and
for one being partially wet ($u_w/\varepsilon=3.5$). A summary of the characteristics
of these defects is given in Table~1 in the main text.

%%%%%%%%%%%%%%%%%%%%%%%%%%%%%%%%%%%%%%%%%%%%%%%%%%%%%%%%%%%%%%%%%%%%%

\section{Free energy profiles off coexistence}

%Fig6
\begin{figure}[h]
	\centering
	\includegraphics[width=0.7\textwidth]{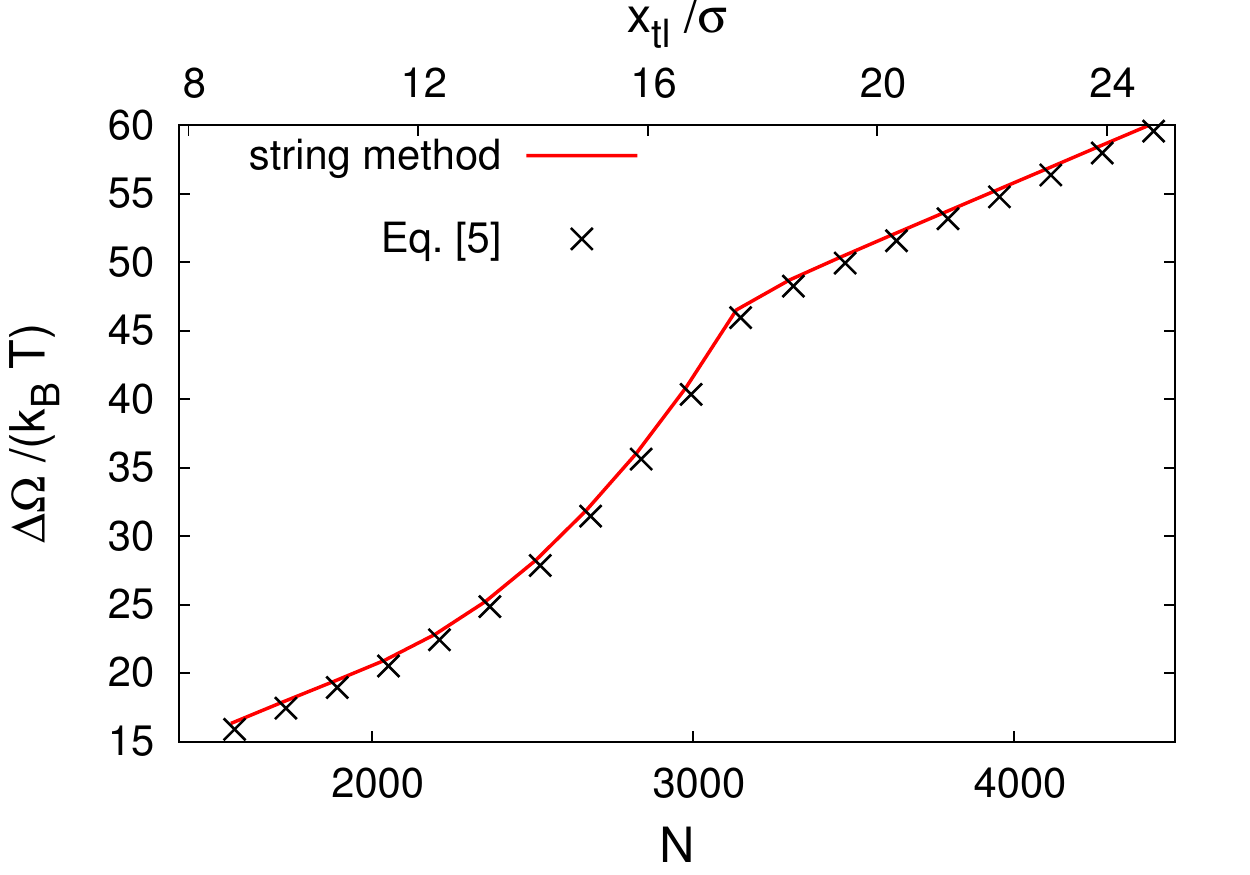}
	\caption{Excess free energy profile $\Delta \Omega$ at $\Delta
	\mu^\ast=-0.01$ for the post defect as computed directly via the string
	method (black crosses) and via Eq.~$[3]$ in the main text.  $N$ is the
	number of particles in the computational box of size $\Delta x \times
	\Delta y \times \Delta z =32.5\sigma \times
	17.5\sigma \times 16\sigma$. 
	As reference, $x_{tl}= N /(L_y L_z (\rho_l -\rho_v))$ is 
	reported, too, on the upper horizontal axis.
	This curve is the same as the upper
	one in Fig.~4A of the main text. 
	Calculations for different $\mu$ 	render similar agreements.
	\label{fig:comparison}}
\end{figure}

Equation~\eqref{eq:shift} of the main text provides a convenient yet approximate
route to compute a free energy profile off coexistence from that at
bulk coexistence. In order to ensure that this is a reliable approximation
within the range of chemical potentials explored here, we have performed
direct string calculations off coexistence and have compared the
resulting free energy profile with that computed via
Eq.~\eqref{eq:CAHrow}.
Figure~\ref{fig:comparison} demonstrates that this approximation is
indeed reliable.

%%%%%%%%%%%%%%%%%%%%%%%%%%%%%%%%%%%%%%%%%%%%%%%%%%%%%%%%%%%%%%%%%%%%%

\section{Macroscopic description of the liquid front}

%Fig7
\begin{figure}[h]
        \centering
	\includegraphics[width=0.90\textwidth]{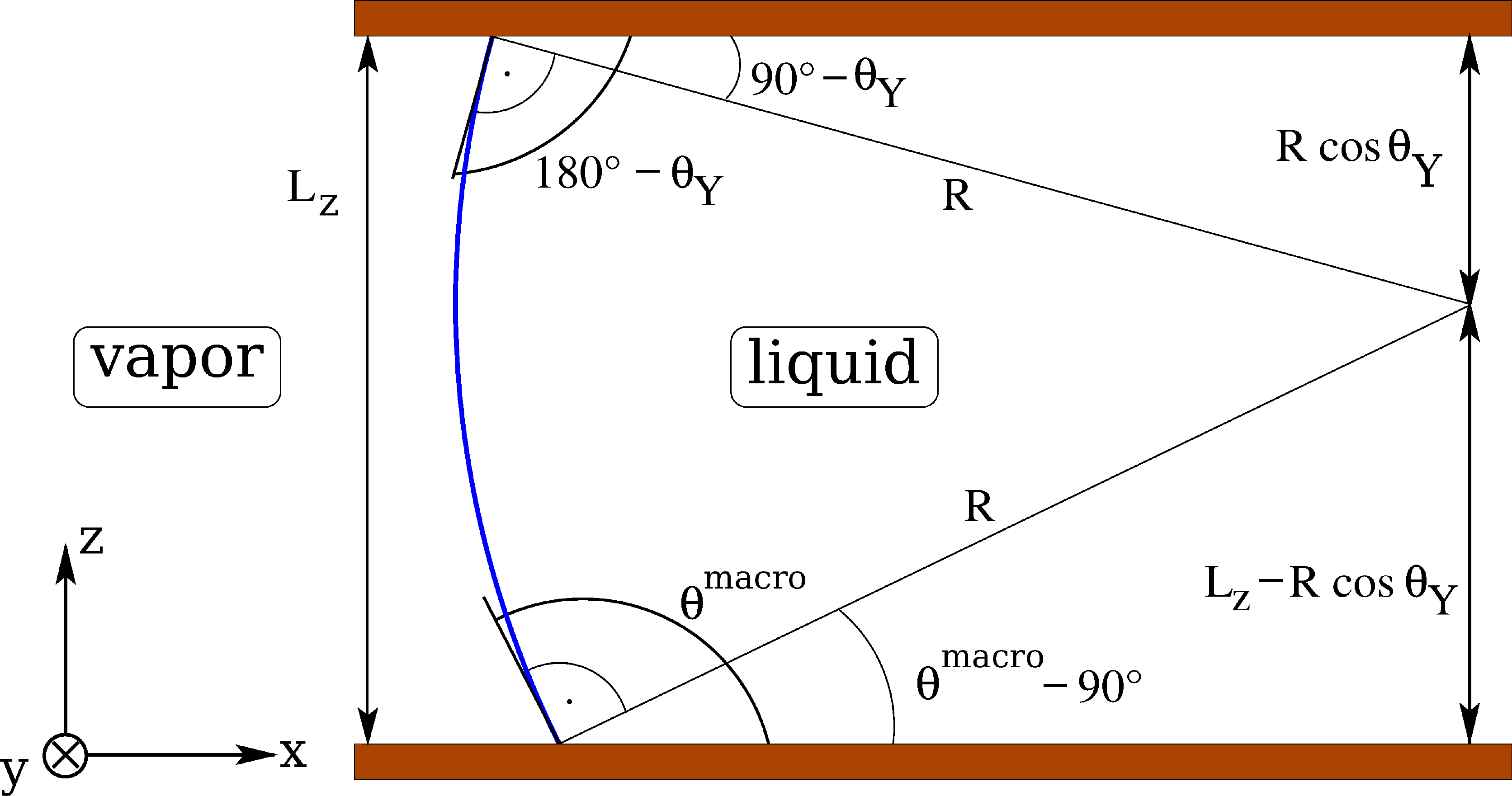}
	\caption{Macroscopic description of a driven liquid front in a
	slit pore of width $L_z$. The cut is perpendicular to the $y$	direction
	defined by the lined-up defects.  The liquid-vapor interface is
	indicated by the blue portion of a circle with radius $R$. The
	prescribed contact angle at the upper wall is $180^\circ - \theta_Y$.
	The macroscopic contact angle $\theta^\mathrm{macro}$ follows from
	geometry.  
        \label{fig:macroscopic}}
\end{figure}

In order to illustrate explicitly the construction of defining the macroscopic 
contact angle, here we provide in addition an alternative route to
Eq.~\eqref{eq:CAHrow} of the main text.
To this end we analyze the same setup as used in our microscopic computations with
the only difference that we now consider a macroscopically large width 
$L_z$ of the channel, which de facto cannot be realized within an entirely
microscopic calculation. In a sufficiently wide channel, large portions of the
liquid-vapor interface are not directly influenced by the walls and by the 
surface defects; these portions can be described in macroscopic terms. On the other hand,
we may use information from our microscopic computations. We know that for driven
systems there are (meta)stable configurations of the contact line and  
we know within which limited range of $\Delta \mu$ or $\Delta p$ they occur. The
portions of the liquid-vapor interface away from the walls attain the macroscopic
shape which is cylindrical for the chosen channel geometry. The radius of curvature
$R = \gamma_{lv}/\Delta p$ (Laplace's law) follows from the pressure difference
$\Delta p$ compatible with the occurrence of (meta)stable configurations, which is known from the 
microscopic computations for any value of $L_z$ via the extrapolation scheme 
discussed above. (In narrow channels, as used in our microscopic computations,
the liquid-vapor interfaces are intrinsically curved as well, but it is difficult to distinguish 
wall and defect-induced curvatures from the one induced by a pressure difference,
because there is no scale separation.) Furthermore, we have prescribed a contact angle close to
$180^\circ - \theta_Y$ on the upper wall which is complementary to Young's angle $\theta_Y$ 
on the defect free part of the lower wall; this specific choice was made in order
to ensure that the liquid-vapor interface is planar in the absence of a driving
pressure difference and in the absence of defects.  
With the knowledge of the curvature radius $R$ and the prescribed contact angle on 
the upper wall, the macroscopic contact angle $\theta^\mathrm{macro}$ on the
lower wall, featuring the pinning defect, follows from purely geometrical considerations.

Figure~\ref{fig:macroscopic} illustrates this geometrical
construction. The macroscopic cylindrical liquid-vapor interface is seen 
as the portion of a circle with radius $R$ intersecting the upper wall at the prescribed contact
angle $180^\circ - \theta_Y$ and the lower wall at the macroscopic contact angle 
$\theta^\mathrm{macro}$ given by the geometrical relation 
\begin{equation}
\sin (\theta^\mathrm{macro} - 90^\circ) = - \cos \theta^\mathrm{macro} = 
 \frac{ L_z - R\cos \theta_Y}{R} \, ,   
\label{eq:anglemacro}
\end{equation} 
as can be read off from Fig.~\ref{fig:macroscopic}.     
Equation~\eqref{eq:anglemacro} can be rewritten as 
\begin{equation}
 \cos \theta^\mathrm{macro} - \cos \theta_Y = - \frac{L_z}{R} 
 \text{ .}
\label{eq:anglemacro-B}
\end{equation}
Using Laplace's law ($\Delta p = \gamma_{lv}/R$) one eventually obtains 
\begin{equation}
        \gamma_{lv}(\cos\theta^\mathrm{macro}-\cos\theta_Y) 
= - L_z \Delta p 
\text{ .}
\label{eq:CAHrow-c}
\end{equation}
By expressing $L_z \Delta p$ in terms of the chemical
potential via $\Delta p = (\rho_l -\rho_v) \Delta \mu$  and by using
Eq.~\eqref{eq:CAHrow-a} of the main text, extrapolated
to macroscopic values of $L_z$ in order to relate $\Delta \mu$ with the computed 
microscopic defect force $\partial \Omega_0/\partial x_{tl}$, one recovers
Eq.~\eqref{eq:CAHrow} in the main text.

%%%%%%%%%%%%%%%%%%%%%%%%%%%%%%%%%%%%%%%%%%%%%%%%%%%%%%%%%%%%%%%%%%%%%
\newpage

\section{Videos}
\label{sec:videos}

\renewcommand{\figurename}{Video}

\begin{figure}[h]
	\centering
	\includegraphics[width=7.5cm]{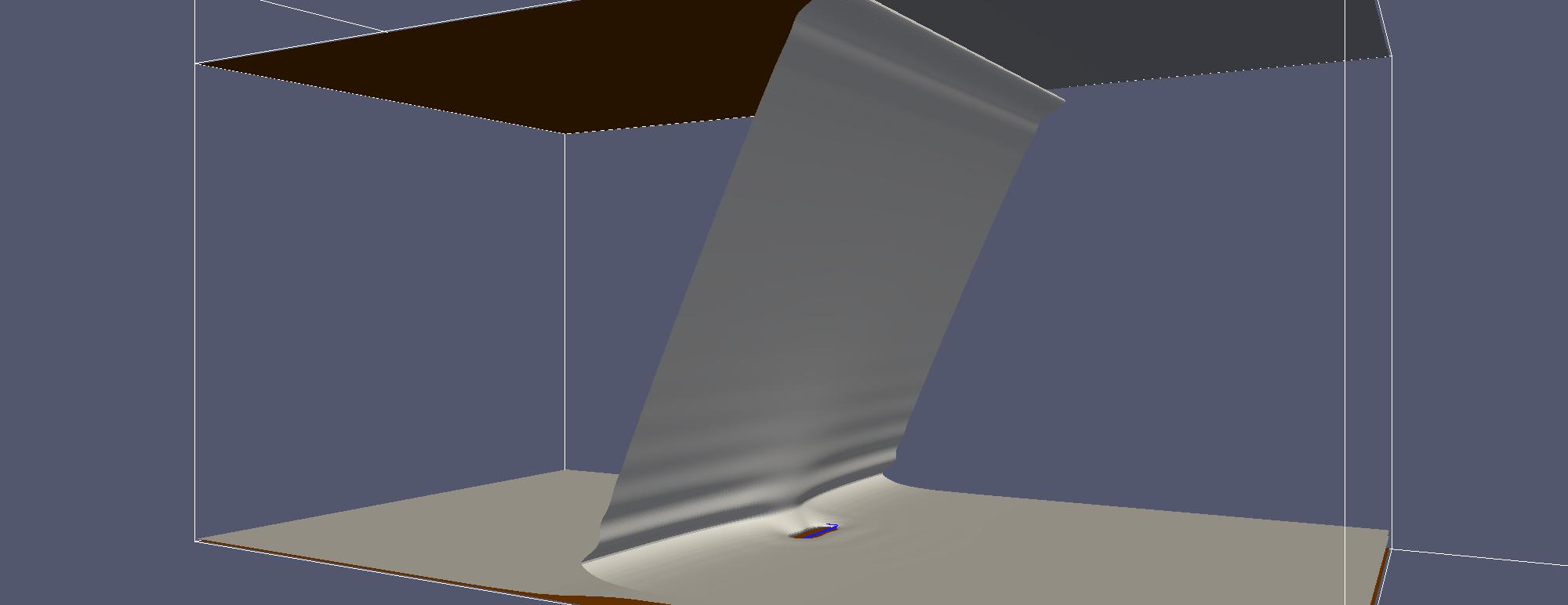}
	\caption{
	Transition path for the advance or retreat of a liquid wedge across a
	lyophobic patch  ($u_w/\varepsilon=2.0$) as computed by using the combined
	DFT and string method. 
	The contour of the 	chemical patch is indicated by the blue lines.
	The iso-potential surface $V(\bm r)=100\,\varepsilon$ (in brown) identifies the upper and lower walls.
	Note that the upper wall is lyophobic whereas the
	lower wall is lyophilic. Therefore on the vapor side a liquidlike film
adsorbed on the wall is visible only at the bottom.
	The video is first played from a rear view through the liquid and then	from a front view	through the vapor.
	\label{vid:lyo}}
\end{figure}

\begin{figure}[h]
	\centering
	\includegraphics[width=7.5cm]{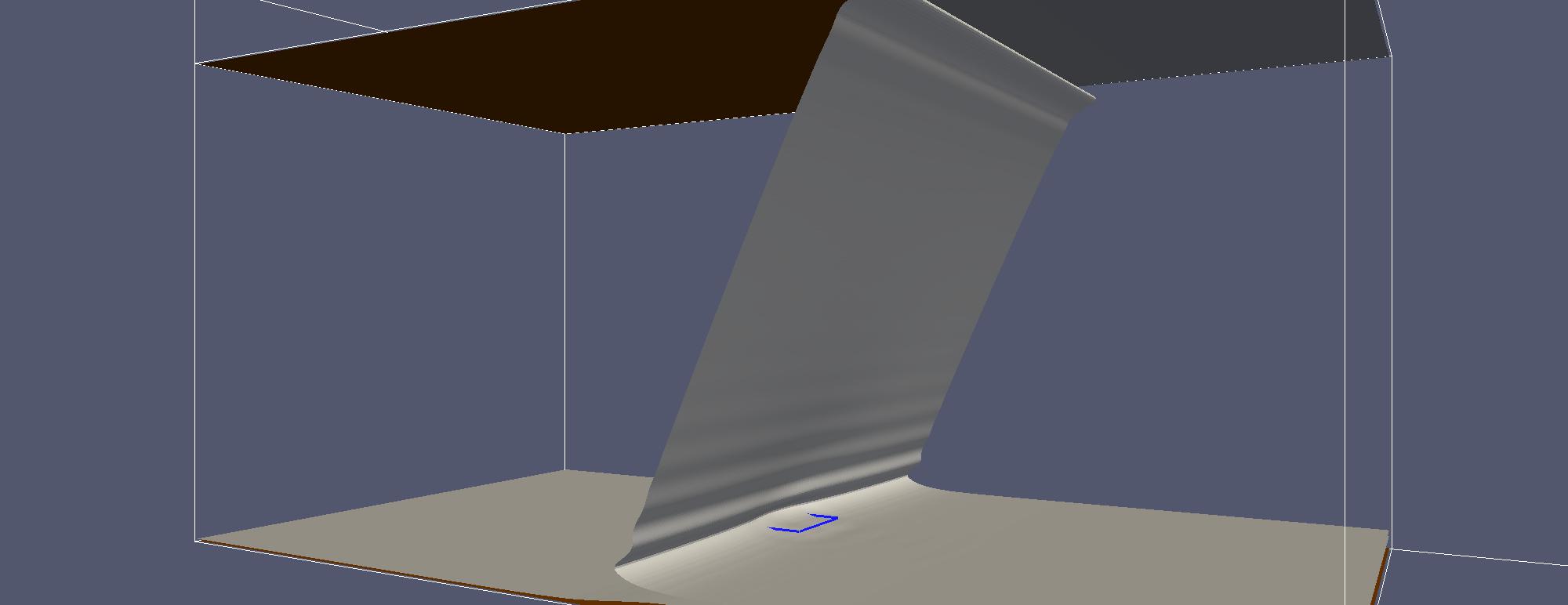}
	\caption{
		Transition path for the advance or retreat of a liquid wedge across a
		partially wet patch ($u_w/\varepsilon=3.5$) as computed
		by using the combined DFT and string
		method.  
		The contour of the 	chemical patch is indicated by the blue lines.
		The iso-potential surface $V(\bm r)=100\,\varepsilon$ (in brown) identifies the upper and lower walls.
	The video is first played from a rear view through the liquid and then	from a front view	through the vapor.
	\label{vid:pwet}}
\end{figure}

\begin{figure}[h]
	\centering
	\includegraphics[width=7.5cm]{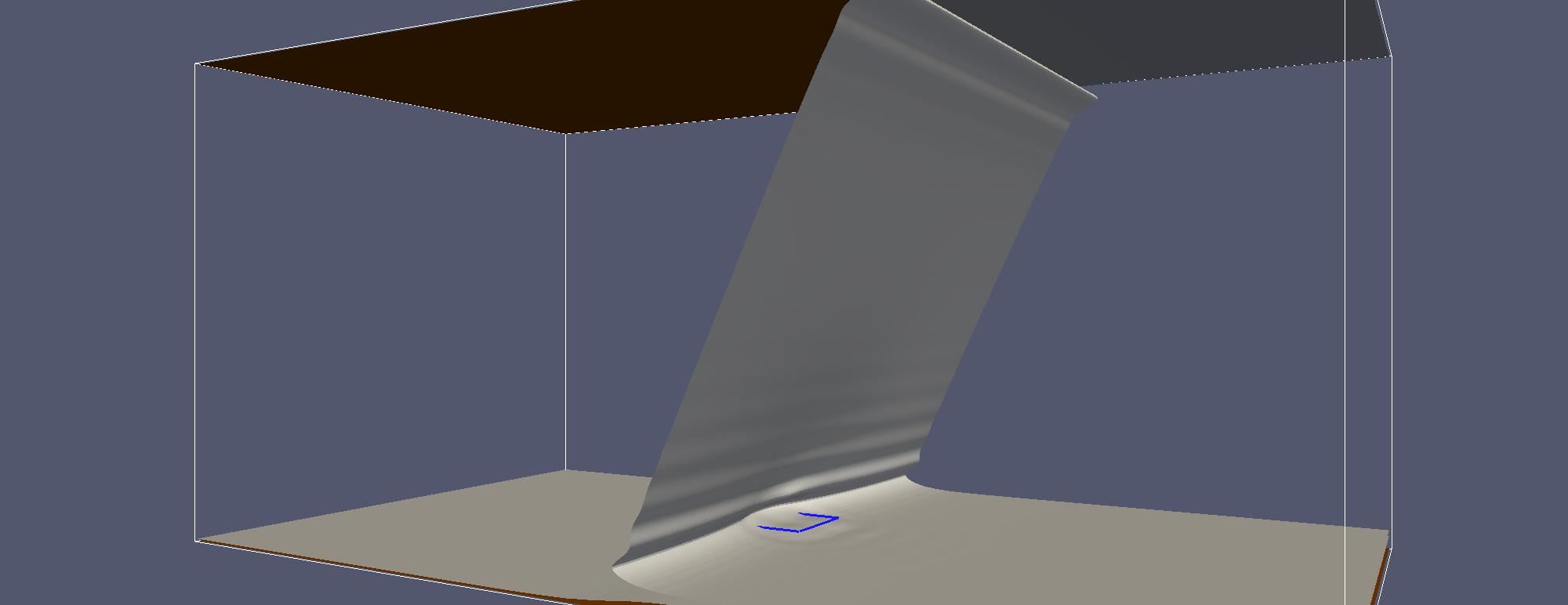}
	\caption{
		Transition path for the advance or retreat of a liquid wedge across a
		completely wet patch ($u_w/\varepsilon=4.0$) as computed
		by using the combined DFT and string
		method. 
		The contour of the 	chemical patch is indicated by the blue lines.
		The iso-potential surface $V(\bm r)=100\,\varepsilon$ (in brown) identifies the upper and lower walls.
	The video is first played from a rear view through the liquid and then	from a front view	through the vapor.
	\label{vid:wet}}
\end{figure}

\begin{figure}[h]
	\centering
	\includegraphics[width=7.5cm]{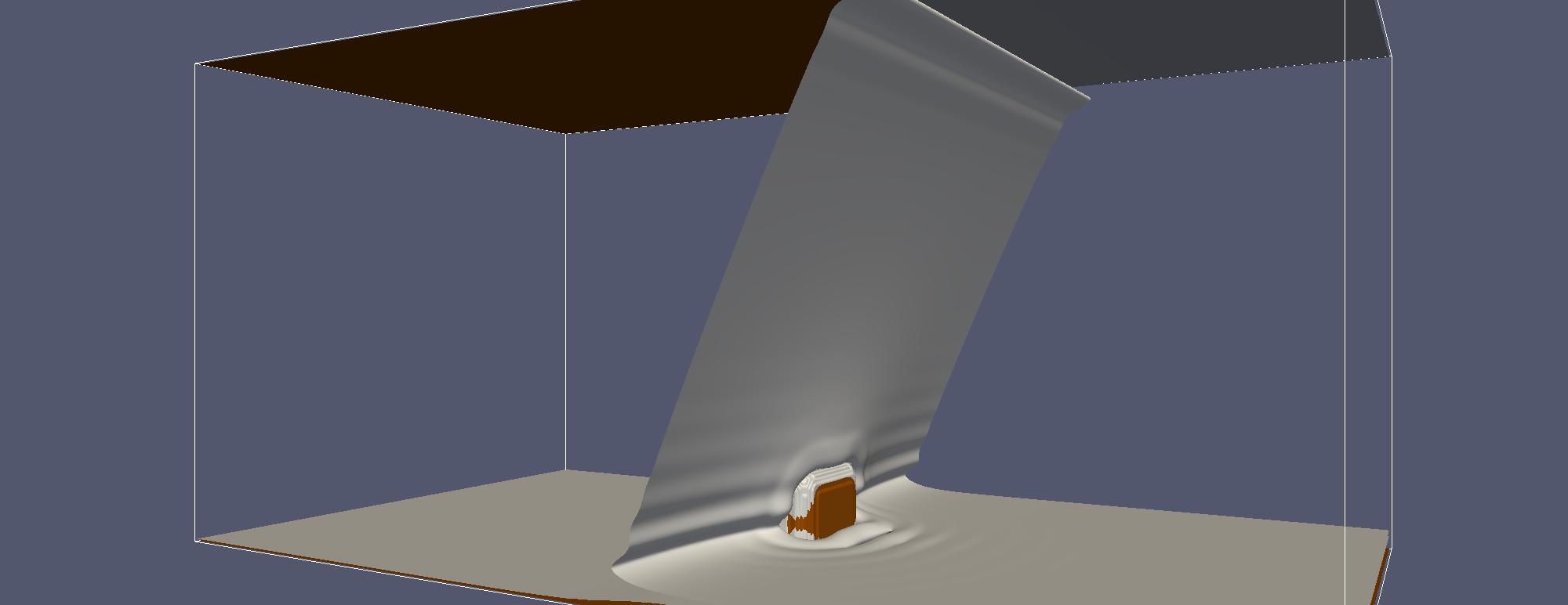}
	\caption{
		Transition path for the advance or retreat of a liquid wedge across a
		cubic post as computed by using the combined DFT and string
		method. 
		The iso-potential surface $V(\bm r)=100\,\varepsilon$ (in brown)
		shows the position of the post and of the upper and lower walls.
	The video is first played from a rear view through the liquid and then	from a front view	through the vapor.
	\label{vid:post}}
\end{figure}

%%%%%%%%%%%%%%%%%%%%%%%%%%%%%%%%%%%%%%%%%%%%%%%%%%%%%%%%%%%%%%%%%%%%%

%%%%%%%%%%%%%%%%%%%%%%%%%%%%%%%%%%%%%%%%%%%%%%%%%%%%%%%%%%%%%%%%%%%%%

%%%%%%%%%%%%%%%%%%%%%%%%%%%%%%%%%%%%%%%%%%%%%%%%%%%%%%%%%%%%%%%%%%%%%

\end{document}